\documentclass[useAMS,usenatbib]{mnras}

\DeclareSymbolFont{cmletters}{OML}{cmm}{m}{it}
\DeclareMathSymbol{v}{\mathalpha}{cmletters}{"76}

\usepackage{tabularx,ragged2e,booktabs,caption}
\usepackage{amsmath}
\usepackage{amssymb}
\usepackage{epsfig}
\usepackage{ifthen}
\usepackage{latexsym}
\usepackage{rotating}
\usepackage{times,epsf}
\usepackage{txfonts}
\usepackage{verbatim}
\usepackage{array}
\usepackage{url}
\usepackage{color}
\usepackage[dvipsnames]{xcolor}
\usepackage[T1]{fontenc}
\usepackage{epstopdf}
\usepackage{ulem}
\usepackage{graphicx}

\newcommand{\be}{\begin{equation}}
\newcommand{\ee}{\end{equation}}

\newcommand{\bea}{\begin{align}}
\newcommand{\eea}{\end{align}}

\newcommand{\ttt}[1]{\texttt{#1}}

\newcommand{\Medd}{\dot M_{\rm Edd}}
\newcommand{\msun}{M_\odot}
\newcommand{\rg}{r_{\rm g}}
\newcommand{\tg}{t_{\rm g}}

\title[Low luminosity accretion flows]{Radiative, two-temperature
  simulations of low luminosity black hole accretion flows
  in general relativity}
\author[A. S\k{a}dowski, M. Wielgus,  R. Narayan, D. Abarca,
J.~C. McKinney, A. Chael]
       {Aleksander S\k{a}dowski$^{1,2}$\thanks{E-mail: asadowsk@mit.edu (AS)}, 
         Maciek Wielgus$^3$\thanks{E-mail: maciek.wielgus@gmail.com (MW)},  Ramesh Narayan$^4$,
         David Abarca$^3$, \newauthor Jonathan C. McKinney$^5$ and
         Andrew Chael$^4$\\
        $^1$ MIT Kavli Institute for Astrophysics and Space Research,
77 Massachusetts Ave, Cambridge, MA 02139, USA\\
$^2$ Einstein Fellow\\
$^3$ Nicolaus Copernicus Astronomical Center, Bartycka 18, Warsaw
00-716, Poland\\
$^4$ Harvard-Smithsonian Center for Astrophysics, 60 Garden St., Cambridge, MA 02134, USA\\
$^5$ Department of Physics and Joint Space-Science Institute,
University of Maryland, College Park, MD 20742\\
}

\begin{document}

\maketitle

\label{firstpage}

\begin{abstract}

We present a numerical method which evolves a two-temperature,
magnetized, radiative, accretion flow around a black hole, within the
framework of general relativistic radiation magnetohydrodynamics. As
implemented in the code KORAL, the gas consists of two sub-components
-- ions and electrons -- which share the same dynamics but experience
independent, relativistically consistent, thermodynamical
evolution. The electrons and ions are heated independently according
to a prescription from the literature for magnetohydrodynamical
turbulent dissipation.
Energy exchange between the particle species via Coulomb collisions is
included. In addition, electrons gain and lose energy and momentum by
absorbing and emitting synchrotron and bremsstrahlung radiation, and
through Compton scattering.
All evolution equations are handled within a fully covariant
framework in the relativistic fixed-metric spacetime of the black
hole.  Numerical results are presented for five models of low
luminosity black hole accretion. In the case of a model with a mass
accretion rate $\dot{M}\sim4\times10^{-8} \Medd$, we find that radiation has
a negligible effect on either the dynamics or the thermodynamics of
the accreting gas. 
In contrast, a model with a larger $\dot{M}\sim 4\times10^{-4} \Medd$
behaves very differently. The accreting gas is much cooler and the
flow is geometrically less thick, though it is not quite a thin
accretion disk.

\end{abstract}


\section{Introduction}
\label{s.introduction}

Black holes are common. Almost every galaxy  in the Universe is believed to
harbor a supermassive black hole (SMBH) at its core, often visible as
an active galactic nucleus, as well as millions of
stellar mass black holes, a few of which accrete gas from 
companion stars in close binary systems and are visible as X-ray
binaries. We now also know that binary black holes are numerous and
that they merge while emitting gravitational waves \citep{ligo+16}.

Because of the compactness of
a black hole,
the gas approaching its horizon liberates a large fraction of its binding
energy. Some of this energy goes into electromagnetic radiation and
makes black hole systems exceptionally bright - black holes in
outbursts are often the brightest X-ray sources in their galaxies, and
the central engines of Active Galactic Nuclei (AGN) can be seen from
across the Universe. The energy and momentum extracted from an AGN
strongly affect not only the inner cluster, but also the dynamics of
the  entire galaxy.

There are three well established regimes of black hole accretion. The
lowest accretion rates result in the lowest densities of gas. Such
accretion flows are optically thin and radiatively inefficient and
correspond to the radio-mode of an AGN and the low-hard state and
quiescent state of X-ray
binaries (analytical models for such conditions were developed
by, e.g., \cite{narayanyi-95} and \cite{blandfordbegelman-99}, see
\cite{yuannarayan-14}, for a review). As the optical depth
grows with increasing accretion rate, the gas cools more efficiently,
and above a critical accretion rate a thin accretion disk is formed \citep{ss73}.
Finally, once the
accretion rate and the corresponding disk luminosity approach the
Eddington values (see next Section), photon trapping and radiative
driving of winds become important and modify the picture
\citep[e.g.,][]{abra88}.

Whenever gas is sufficiently dense, the
electrons and ions equilibrate efficiently because of Coulomb
coupling and their temperatures become equal. 
This fact greatly simplifies 
the physics involved - the plasma can be treated as a single
temperature fluid with mean molecular weight reflecting its chemical
composition.

Optically thin flows are more complicated. The Coulomb interaction is
inefficient and the temperatures of electrons and ions can be very
different. The actual distribution of energy between the species
depends not only on the balance between radiative cooling (acting
on electrons) and Coulomb coupling (bringing electrons and ion temperatures
together), 
but also on how the electrons and ions are heated. The latter
depends on microphysics which cannot be resolved in
magnetohydrodynamical grid-based simulations such as the ones reported in this
paper. Instead one has to make some strong assumptions or adopt
sub-grid prescriptions, which are generally somewhat ad hoc since our understanding of the heating process
is still limited \citep[see][for a review]{yuannarayan-14}.

So far, only accretion flows corresponding to the two limits,
optically thin and thick, have been simulated consistently in the
framework of multidimensional magnetohydrodynamics. Simulations of accretion at the lowest
rates, $\dot M\lesssim 10^{-6} \Medd$, for long have
only been done using the single fluid approximation
\citep[e.g.,][]{hawley-00,tchekh10,narayan+adaf,mtb12,moscibrodzka+14}.

In the case of an optically thin medium,
the radiation field is dynamically unimportant and can be solved for in
the postprocessing stage \citep[e.g.,][]{dexter+10,roman+12,ck+15a}. Because only a
single fluid (mixture of ions and electrons) is evolved in most MHD simulations, one has to choose, somewhat arbitrarily, what the electron
temperature is, which ultimately determines the emitted
radiation. Only recently has this method been improved.
\cite{ressler+15} introduced an independent
second fluid, electrons, and they used a prescription for the heating
of electrons and ions according to a physically
motivated sub-grid model \citep{howes+10}. This approach leads to the
ability to determine electron temperatures with fewer ad hoc prescriptions.

In the case of an
optically thick gas, one does not need to distinguish between electrons and ions   - both have the same temperature because of
efficient Coulomb coupling. However, in such cases, radiation is
dynamically important and has to be solved in parallel with the
evolution of the gas and the magnetic field. Recent developments in global
simulations of accretion flows of this kind (corresponding to
accretion rates $\dot M\gtrsim 10^{-3} \Medd$) include the pioneering
works of \cite{ohsuga+09} and \cite{ohsugamineshige-11}, as well as
\cite{sadowski+koral,sadowski+koral2,mckinney+harmrad,fragile+14}, who
applied general relativistic M1 closure, \cite{jiang+14} who directly
solved the radiation transfer equation in a Newtonian potential,
and \cite{ryan+15} who solved the equations of general
  relativistic 
radiation MHD using a direct Monte Carlo solution of the frequency-dependent radiative transport equation.

None of these methods are appropriate in the regime in between,
$10^{-6}\Medd\lesssim \dot M\lesssim 10^{-3}\Medd$.  At such accretion
rates, radiative cooling starts to
affect the gas temperature, but the cooling is not as efficient as in
the optically thick case. Coulomb coupling brings the electron and
ion temperatures towards each other, but does not yet equalize them.

The intermediate regime of accretion rates is relevant both for
supermassive and stellar mass BHs. Many bright AGN are in this state,
with M87 \citep[e.g.,][]{biretta+91} being a prominent
example. X-ray binaries spend significant fraction of their lifetime at such low 
accretion rates. During the hard-to-soft transition, the accretion rate
increases and the system often enters the high-soft, optically thick,
state. Understanding the nature of this transition, and the physical
processes behind the related phenomena, e.g., iron line reflection and
quasi periodic oscillations, will not be possible without detailed
numerical modelling of accretion flows in the intermediate regime.

In this paper we introduce a hybrid algorithm which evolves the gas and
radiation fields together in parallel, and accounts for their exchange of energy and momentum.
At the same time, the algorithm evolves electrons and ions as
two separate fluids affected by radiative cooling, Coulomb coupling, and viscous
heating. This algorithm has been implemented in a general
relativistic radiative magnetohydrodynamic (GRRMHD) code \texttt{KORAL}. We also present and adopt sophisticated
synchrotron and bremsstrahlung grey opacities. Such a~combination allows us, for the first time, to 
consistently simulate accretion flows in the intermediate regime.

The paper is organized as follows. In Section~\ref{s.physics} we give
all the essential equations describing conservation of mass, energy
and momentum, electron and ion entropy evolution, viscous heating,
Coulomb coupling, and the radiative coupling term. In particular, in
Section~\ref{sec:opacities} we give an expanded form of the
opacities. In Section~\ref{s.method} we describe in detail the
numerical algorithm, which we test in Section~\ref{s.tests}. In
Section~\ref{s.low} we present and discuss simulations of low
luminosity accretion flows. Finally, in Section~\ref{s.discussion} we
discuss various caveats, and we conclude in Section~\ref{s.summary}.

\subsection{Units and formalism}

In this work we adopt the following definition
for the Eddington mass accretion rate,
\be
\label{e.medd}
\Medd = \frac{L_{\rm Edd}}{\eta c^2},
\ee
where $L_{\rm Edd}=4\pi GMm_{\rm p} c/\sigma_{\rm T}=1.25 \times 10^{38}  M/M_{\odot}\,\rm erg/s$ is the 
Eddington luminosity for a~BH of mass $M$, and $\eta$ is the radiative efficiency of a~thin
disk around a~black hole with a~given spin $a_* \equiv a/M$. For a zero
BH spin, the case considered in this paper, $\eta\approx 0.057$ and
$\Medd = 2.48 \times 10^{18}M/M_{\odot}  \,\rm g/s$.
Hereafter, we use the
gravitational radius $r_{\rm g}=GM/c^2$ as the unit of length, and
$t_{\rm g}=r_{\rm g}/c$
as the unit of time.

\section{Physics}
\label{s.physics}

\subsection{Standard set of equations}

The conservation laws for gas density, its energy and momentum,
radiation energy and momentum, and photon number can be written in covariant form,
\begin{align}
(\rho u^\mu)_{;\mu}&= 0 ,\label{eq.cons1} \\
(T^\mu_{\ \nu})_{;\mu}&= G_\nu,\label{eq.cons2}\\
(R^\mu_{\ \nu})_{;\mu}&= -G_\nu,\label{eq.cons3}\\
(n u^\mu)_{;\mu}&= \dot n.  \label{eq.cons4}
\end{align}
where $\rho$ is the gas
density in the comoving fluid frame, $u^\mu$ is the gas four-velocity,
$T^\mu_\nu$ is the
MHD stress-energy tensor,
\be\label{eq.tmunu}
T^\mu_{\ \nu} = (\rho+u_{\rm int}+p+b^2)u^\mu u_\nu + (p+\frac12b^2)\delta^\mu_{\ \nu}-b^\mu b_\nu,
\ee 
with $u_{\rm int}$ and $p=(\gamma_{\rm int}-1)u_{\rm int}$ representing the internal energy density and pressure of the 
gas in the comoving frame with adiabatic index, $\gamma_{\rm int}$ (as
explained in 
Appendix~\ref{ap.entropy}, we consider two different
kinds of adiabatic index, $\gamma_{\rm int}$ and $\gamma_{\rm CV}$,
one for the internal energy and one for the specific heat), $R^\mu_\nu$ stands for the radiative stress-energy
tensor  and $n$ for the photon number density, and $b^\mu$ is the
magnetic field 4-vector which is evolved following the ideal-MHD induction equation
\citep{gammie03}. 

The radiative stress-energy tensor is obtained from the evolved
radiative primitives, i.e., the radiative rest-frame energy density,
$E_{\rm R}$, and its four-velocity, $u^\mu_{\rm R}$, following the M1
closure scheme
 modified by the addition of radiative viscosity \citep[see][for details]{sadowski+koral,sadowski+dynamo}.

The interaction between the gas and the radiation, i.e., the transfer
of energy and momentum, is described by 
the radiation four-force density $G_\nu$.
The opposite signs of this quantity on the right hand sides
of Eqs.~\ref{eq.cons2} and \ref{eq.cons3} reflect the fact
that the gas-radiation interaction is conservative, i.e., it transfers energy
and momentum between gas and radiation. The detailed form of the
four-force density is given and discussed in Section~\ref{s.gt} and
the relevant opacities are presented in Section~\ref{sec:opacities}
and Appendix~\ref{ap.entropy}.

\subsection{Evolution of ions and electrons}

The set of equations (Eqs.~\ref{eq.cons1}-\ref{eq.cons4}) given above 
describes the evolution of the \textit{total} gas energy density and
momentum. The emission properties of the gas, however, depend mostly
on the electrons. 
It is
reasonable to assume that both electrons and ions move with the same bulk
velocity, i.e., there is no need to consider their momenta independently. However, it is necessary to follow the
energy content of the two species. In principle, it is enough to
follow the energy of only one of
the species because the total energy is evolved already in the
standard set of equations. However, for the sake of symmetry and
accuracy, and
keeping in mind the very moderate cost of evolving another equation, we
decided to follow the energy content in the electrons and ions independently.

We adopt an approach similar to the one taken by \cite{ressler+15},
i.e., we evolve ions and electrons adiabatically by conserving their
entropy, and apply the non-adiabatic changes to their energy as source
terms. 

The entropy equations of the two species take the form,
\begin{align}
T_{\rm e}(n_{\rm e} s_{\rm e} u^\mu)_{;\mu}&=\delta_{\rm e}
q^{\rm v}+q^{\rm C}+G_t \label{e.entr1}\\
T_{\rm i}(n_{\rm i} s_{\rm i} u^\mu)_{;\mu}&=(1-\delta_{\rm e}) q^{\rm
  v}-q^{\rm C}, \label{e.entr2}
\end{align}
where $T_{\rm e}$ and $T_{\rm i}$ are the temperatures of ions and
electrons, respectively, $n_{\rm e}$ and $n_{\rm i}$ are the number densities,
$s_{\rm e}$ and $s_{\rm i}$ are the entropies per particle (which are given in
Section~\ref{s.entropy}), $q^{\rm v}$ is the rate of dissipative
heating, a fraction of which ($\delta_{\rm e}$) is applied to electrons, and
the remaining part ($1-\delta_{\rm e}$) to ions (Section~\ref{s.viscous}),
$q^{\rm C}$ is the Coulomb coupling rate which describes energy
exchange between electrons and ions (Section~\ref{s.coulomb}), and
finally, $G_t$ is the radiative heating/cooling rate which affects
only electrons (Section~\ref{s.gt}). Setting the right hand side of
Eqs.~\ref{e.entr1} and \ref{e.entr2} to zero corresponds to adabatic
evolution of the species.

The electron and ion number densities are,
\begin{align}
 n_{\rm e,i}& = \dfrac{\rho}{\mu_{\rm e,i}m_{\rm p}},
\end{align}
where $m_{\rm p}$ is the proton mass and $\mu_{\rm e}$ and $\mu_{\rm i}$
 are the mean molecular weights of electrons and ions, respectively,
 which, for zero metallicity, are given by,
\begin{align}
 \mu_{\mathrm{e}} &= \dfrac{2}{1+X},\\
 \mu_{\mathrm{i}} &= \dfrac{4}{4X+Y},
\end{align}
where $X$ and $Y$ denote the relative mass abundances of hydrogen and
helium.
In the simulations presented in this paper we exclusively use
$X= 1$, $Y = 0$.

\subsubsection{Entropy, Adiabatic Index and Temperature}
\label{s.entropy}

In this work we use the following approximate 
relativistic entropy formula derived in Appendix~\ref{ap.entropy},
\begin{equation}
\label{eq.s}
s_{\rm e,i} = k \ln \left[ \frac{\theta_{\rm e,i}^{3/2} \left(\theta_{\rm e,i} + \frac{2}{5}\right)^{3/2}}{\rho_{\rm e,i}}
\right],
\end{equation}
where $\theta_{\rm e,i}=kT_{\rm e,i}/m_{\rm e,i}c^2$ are the
dimensionless temperatures of electrons and ions and $m_{\rm e,i}$ are
the particle masses. 


Inverting Equation~(\ref{eq.s}), we can solve for the temperature
$\theta$ as a~function of the density and entropy per particle, 
\be
\label{e.TfromS3}
\theta_{\rm e,i}=\frac{1}{5}\left(\sqrt{1+25 \left[\rho_{\rm e,i}
    \exp\left(\frac{s_{\rm e,i}}{k}\right) \right]^{2/3}}-1\right).
\ee

\subsubsection{Viscous heating}
\label{s.viscous}

We measure the amount of the viscous heating affecting the gas, i.e.,
electrons and ions altogether, by comparing the non-adiabatic
evolution according to the energy equation of the total gas (Eq.~\ref{eq.cons2}), with the adiabatic
evolution of electrons and ions (Eqs.~\ref{e.entr1} \& \ref{e.entr2}
with right hand sides set to zero). The former is expected to lead to a
larger energy density than the latter, and the difference comes solely
from viscous dissipation.

Therefore, the viscous heating rate in the comoving frame of the gas, $\widehat{ q^{\rm v}}$, is given by,
\be
\label{e.qv}
\widehat {q^{\rm v}}=\frac{u_{\rm int} - u_{\rm e,adiab.} - u_{\rm i,adiab.}}{\Delta \tau},
\ee
where $u_{\rm int}$ is the internal energy density of the total gas at the end of the
explicit (convective) operator (see Section~\ref{s.implementation})
obtained by evolving the time (energy) component of Eq.~\ref{eq.cons2}
over comoving frame time step $\Delta \tau$,
and $u_{\rm e,adiab.}$  and $u_{\rm i,adiab.}$ correspond to the
internal energies carried by electrons and ions, respectively, measured once again at the end
of the explicit operator and
obtained through the adiabatic evolution of their respective
entropies.

The (adiabatic) energy densities of the electrons and ions are
obtained from their entropy
densities by first calculating their respective temperatures through
Eq.~\ref{e.TfromS3}. Knowing the temperatures, we may write down the
electron and ion pressures,
\be
\label{e.pei}
p_{\rm e,i}=k\frac{\rho}{\mu_{\rm e,i}m_{\rm p}}T_{\rm e,i},
\ee
and obtain the corresponding energy densities through,
\be
\label{e.uei}
u_{\rm (e,i),adiab.}=\frac{p_{\rm e,i}}{\gamma_{\rm int\,\rm e,i}-1},
\ee
where $\gamma_{\rm int\,e,i}$ are the electron and ion adiabatic indices
corresponding to their temperatures (Appendix~\ref{ap.entropy}),
\begin{equation}
\gamma_{\rm int\,\rm e,i} \approx \frac{10+20\theta_{\rm e,i}}{6+15\theta_{\rm e,i}}.
\label{eq:gammaint0}
\end{equation}

The total viscous heating rate, $q^{\rm v}$ (Eq.~\ref{e.qv}) is finally distributed
between electrons and ions. The fraction of heating going into
electrons depends on the microscopic properties of a collisionless
plasma. In this work we follow \cite{ressler+15} and use the fitting formula of \cite{howes+10}
for the fraction, $\delta_{\rm e}$, of the total heating that goes
  into electrons,
\begin{equation}
\label{e.deltae}
\delta_{\rm e}= \frac{1}{1 + f(T_{\rm e}, T_{\rm i}, \beta)},
\end{equation}
\begin{equation}
\label{e.deltae2}
f(T_{\rm e}, T_{\rm i}, \beta) = c_1 \frac{c_2^2 + \beta^{2 - 0.2 \log_{10}(T_{\rm i}/T_{\rm e})} }{c_3^2 + \beta^{2 - 0.2 \log_{10}(T_{\rm i}/T_{\rm e}) } }\sqrt{\frac{m_{\rm i} T_{\rm i}}{m_{\rm e} T_{\rm e}}} \exp(-1/ \beta) \ ,
\end{equation}
where $\beta$ is the gas to magnetic pressure ratio,
\(c_1 = 0.92\), \(c_2 = 1.6T_{\rm e}/T_{\rm i}\) and \(c_3 = 18 +5 \log_{10}(T_{\rm i}/T_{\rm e})\) 
if \(T_{\rm i} > T_{\rm e}\), while we have \(c_2 = 1.2T_{\rm
  e}/T_{\rm i}\) and \(c_3 = 18\) for \(T_{\rm i} < T_{\rm e}\).

\subsubsection{Coulomb coupling}
\label{s.coulomb}

The ions and electrons exchange energy kinetically, through collisions. This effect is described by the comoving frame Coulomb coupling rate, which equals \citep{stepneyguilbert-83},
\begin{align}
\widehat{q^{\rm C}} &=  \frac{3}{2}   \frac{m_{\rm e}}{m_{\rm i}} \overline{n} n_{\rm e} \ln \Lambda
\frac{c k \sigma_T ( T_{\rm i} - T_{\rm e})}{K_2(1/ \theta_{\rm i}) K_2(1 / \theta_{\rm e})} \times \nonumber \\
&\left[ \frac{2( \theta_{\rm e} + \theta_{\rm i})^2 + 1}{\theta_{\rm e} + \theta_{\rm i}} K_1(1/ \theta_{\rm m}) + 2 K_0 (1 / \theta_{\rm m}) \right] \  {\rm \left[\frac{erg}{ cm ^{3} s}\right]} ,
\label{eq:CoulombC}
\end{align}
with
$\theta_{\rm m} = \left(1/ \theta_{\rm e} + 1/ \theta_{\rm i} \right)^{-1}$, $m_{\rm i} =
m_{\rm p}(X + 4Y)$. $K_i$ denotes the modified Bessel function of the
$i$-th order and $\ln \Lambda \approx 20$ is the Coulomb logarithm. Number density $\overline{n}$ is defined as
\be
\overline{n} = \sum_j \left( n_{\rm i} \right)_j Z_j^2=(X+Y)\rho / m_p=\rho/m_p,
\ee
where $Z_j$ is the charge of the $j$-th species of ions, and the last
equality holds for a pure mixture of hydrogen and helium. To avoid the computational problems related to evaluation of the Bessel functions for very large arguments, in the low temperature limit we continuously match them with the asymptotic formula
\be 
K_{0 ,1 , 2}\left(1/\theta \right)\approx  \exp\left(-1 / \theta \right) \sqrt{\frac{\pi \, \theta}{2}} .
\ee

\subsection{Interaction of gas and radiation}
\label{s.gt}

\subsubsection{Temperatures and the equation of state}
\label{s.temps}

Before we discuss in detail the interaction of gas and radiation we
reiterate the various temperatures evolved in the code.

(i) $T_{\rm g}$ is the effective temperature of the gas (mixture of the electrons and
ions). It is given through the ideal gas equation of state, 
\be
\label{e.Tgas}
p_{\rm g}=(\gamma_{\rm gas}-1)u_{\rm int}=k\frac{\rho}{\mu m_{\rm p}}T_{\rm g},
\ee
where $p_{\rm g}$ and $u_{\rm int}$ are the gas pressure and internal
energy density, respectively, $\gamma_{\rm gas}$ is the effective adiabatic
index of the electron and ion mixture (see below), and $\mu$ is the mean molecular
weight,
\begin{equation}
\mu = \left(\frac{1}{\mu_{\rm i}} + \frac{1}{\mu_{\rm e}} \right)^{-1} = \frac{4}{6X + Y + 2}.
\end{equation}

(ii) $T_{\rm e}$ and $T_{\rm i}$ are the ion and electron
temperatures, respectively. They are obtained from the evolved entropy
density through Eq.~\ref{e.TfromS3} and satisfy the respective
equations of state,
\be
\label{e.Tei2}
p_{\rm e,i}=(\gamma_{\rm int\,e,i}-1)u_{\rm e,i}=k\frac{\rho}{\mu_{\rm e,i} m_{\rm p}}T_{\rm e,i}.
\ee
The electron and ion pressures, $p_{\rm e}$ and $p_{\rm i}$, satisfy,
\be
p_{\rm g}=p_{\rm e} + p_{\rm i}.
\ee
The temperatures must therefore satisfy,
\begin{equation}
T_{\rm g} = \mu\left(\frac{T_{\rm i}}{\mu_{\rm i}} + \frac{T_{\rm e}}{\mu_{\rm e}} \right).
\label{e.Tgmix}
\end{equation}
The effective adiabatic index for the electron and ion mixture is
thus,
\be
\gamma_{\rm gas}=1+\frac{(\gamma_{\rm int\, e}-1) (\gamma_{\rm int\,i}-1)
(T_{\rm i}/T_{\rm e}+\mu_{\rm i}/\mu_{\rm e})}{(T_{\rm i}/T_{\rm e})
(\gamma_{\rm int\, e}-1) 
+(\mu_{\rm i}/\mu_{\rm e}) (\gamma_{\rm int\, i}-1)}
\label{e.gammagas}
\ee


(iii) $T_{\rm r}$ is the characteristic temperature of radiation,
  calculated from the radiation energy density and photon number
  density. It roughly corresponds to
the frequency at which the electromagnetic spectrum of radiation peaks
in the fluid frame at a~given time and location.

We make
the simplifying assumption that the spectral energy density of radiation $\widehat{E}_\nu(\nu , T_{\rm r})$
corresponds to a diluted black body (grey body) spectrum, i.e.,
\be 
\widehat{E}_\nu(\nu , T_{\rm r}) = \phi \widehat{B}_\nu(\nu , T_{\rm r}) = \phi \frac{2 h \nu^3}{c^2} \frac{1}{\exp(h \nu / k T_{\rm r}) -1},
\ee
for a non-negative constant coefficient $\phi \le 1$. KORAL tracks the total radiation energy
density $\widehat{E}$, and photon density $\widehat n$ 
\citep[see][]{sadowski+compt}, from which  we obtain,
\begin{equation}
T_{\rm r} = \frac{\widehat{E}}{2.7012 k \widehat{n}}  \; \; \;,  \; \;
\; \phi = \frac{\widehat {E} c}{4 \sigma T_{\rm r}^4} .
\label{eq:Tr}
\end{equation}

\subsubsection{The coupling term}
\label{s.Gi}

Gas and radiation exchange energy and momentum through absorption and
scattering. This interaction is described by the source term - the
radiative four-force, $G_\mu$, which couples
gas and radiation (Eqs.~\ref{eq.cons2}-\ref{eq.cons3}). In the comoving
frame of the gas it takes the following form,
\be\label{eq.Gff2}
\widehat G= \rho
\left[ \begin{array}{c}
 \kappa_{\rm P, a} \, \widehat E  -4\pi \kappa_{\rm P, e} \, \widehat
 B+ \widehat G_{t,\rm Compt}\\
 (\kappa_{\rm R} +\kappa_{\rm es})  \widehat F^i 
\end{array} \right].
\ee
Here $\widehat B=\sigma T_{\rm e}^4/\pi$ is the black body radiance of
the electrons, $\widehat F^i$ is the radiation flux in the fluid
frame, and we sum the Rosseland and scattering opacities for
simplicity (in
principle, the Rosseland mean opacity should be a single integral over all opacities, including the scattering term).
In this work we account for bremsstrahlung and synchrotron
opacities, hence, each opacity coefficient $ \kappa_{\rm P, a}$ (energy mean absorption), $ \kappa_{\rm P, e}$ (energy mean emission) and $\kappa_{\rm R, a}$ (Rosseland mean absorption) consists of bremsstrahlung and synchrotron components, i.e.,
\begin{align}
\kappa_{\rm P, e} &=\kappa^{\rm (ff)}_{\rm P, e} + \kappa^{\rm (sy)}_{\rm P, e},\\
\kappa_{\rm P, a} &=\kappa^{\rm (ff)}_{\rm P, a} + \kappa^{\rm (sy)}_{\rm P, a},\\
\kappa_{\rm R} &=\kappa^{\rm (ff)}_{\rm R} + \kappa^{\rm (sy)}_{\rm R}.
\end{align}
These opacities are frequency-averaged, effective, gray
opacities. There is no unique way of defining or computing these
quantities since they depend on the precise frequency dependence of
the radiation field. \cite{mihalasbook} recommend using the Planck
mean opacity for the energy equation (time component of $\widehat{G}$)
and the Rosseland mean opacity for the momentum equations (spatial
components). This is the approach we take.

The electron scattering opacity, $\kappa_{\rm es}$, is responsible
both for momentum and energy transfer. The latter is described through
the Comptonization component, $\widehat G^0_{\mathrm{Compt}}$
\citep{sadowski+compt},
\begin{align}
\widehat{G^0}_{\mathrm{Compt}} &=
- \kappa_{\mathrm es}\rho\widehat{E} \left[\frac{4k(T_{\mathrm e}-T_{\mathrm{r}})}{m_{\mathrm e}c^2}\right] \times \label{CompG02} \\
& \left[1+3.683 \left(\frac{kT_{\mathrm e}}{m_{\mathrm e}c^2}\right) 
+4 \left(\frac{kT_{\mathrm e}}{m_{\mathrm e}c^2}\right)^2\right]
\left[1+\left(\frac{kT_{\mathrm e}}{m_{\mathrm e}c^2}\right)\right]^{-1} \! . \nonumber
\end{align}

\subsection{Opacities}
\label{sec:opacities}
In our opacity model we make several improvements over the previous
global GRRMHD simulations such as those reported in
\citet{dibi+12,fragile+14, mckinney+harmrad,sadowski+compt}.  As
indicated by Eq. \ref{eq.Gff2}, we distinguish between the energy mean
(Planck mean) opacities used for the energy term $\widehat{G}^0$ and
the Rosseland mean opacities that are used in the momentum equations.
However, as discussed in \cite{mihalasbook}, there is
  no single gray opacity that is accurate in every opacity regime.  We
  use the energy mean opacity for the energy equation  to capture proper energy balance and the
  Rosseland mean opacity for the momentum equation in order to capture the proper flux in the diffusion
  limit.
We further distinguish between the emission opacities from
the absorption opacities. The emission opacities are calculated for the local electron temperature $T_{\rm e}$, while absorption opacities account for the actual radiation
temperature $T_{\rm r} \neq T_{\rm e}$. In this section we only give
the relevant formulas. More detailed discussion and derivation is given in the Appendix \ref{ap.opacities}.

\subsubsection{Free-free}
We take the bremsstrahlung energy mean emission opacity to be,
\be
\label{eq:OpacEnergyEmiFF}
\kappa^{\rm (ff)}_{\rm P, e} \, \rho = 6.2 \times 10^{-24} \overline{n} n_{\rm e}  T_{\rm e}^{-7/2} \,  \overline{g} \, R(T_{\rm e}) \ {\rm [ cm^{-1} ] } ,
\ee
where we assume a Gaunt factor $\overline{g} = 1.2$ and employ a relativistic
correction \citep{rybicki},
\be
R(T_{\rm e}) = 1 + 4.4 \times 10^{-10}T_{\rm e} .
\label{eq:correctionff}
\ee
The absorption energy mean opacity is given in the following form
\be 
\label{eq:OpacEnergyAbsFF}
\kappa^{\rm (ff)}_{\rm P, a } = \kappa^{\rm (ff)}_{\rm P, e} \xi^{-3} \cdot 1.047 \ln  \left(1 \! + \! 1.6 \xi \right)  ,
\ee
where $\xi = T_{\rm r}/T_{\rm e}$. The absorption opacity, Eq. \ref{eq:OpacEnergyAbsFF}, reduces to the emission opacity, Eq. \ref{eq:OpacEnergyEmiFF}, for $\xi = 1$.\\
Another fit is provided for the Rosseland mean opacity,
\be 
\kappa^{\rm (ff)}_{\rm R } = \kappa^{\rm (ff)}_{\rm P, e} \xi^{-3} \cdot 14.12  f_{\rm R}(\xi) ,
\ee
where
\be 
f_{\rm R}(\xi) = \left(432.7 - 106.8 \xi^{-3/5} + 43.17 \xi^{-4/5} + 57.88 \xi^{-1}  \right)^{-1} .
\ee
When $T_{\rm r} = T_{\rm e}$, the Rosseland mean opacity is about 30 times less than the energy mean opacity. Corresponding opacities are derived in the Appendix \ref{ap.opacities}, Eqs. \ref{eq:calcFFemisOpac}-\ref{eq:calcFFRossOpac}.

 \begin{figure}
  \includegraphics[width=1.\columnwidth]{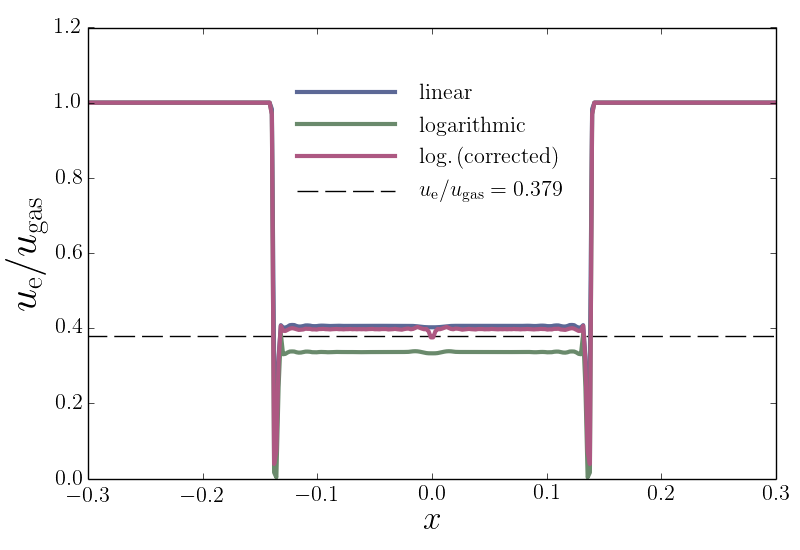}
\caption{Profiles of the ratio of energy densities in electrons
    and gas for the 1D shock test described in
    Section~\ref{s.shock}. The corrected logarithm approach
    (eq.~\ref{eq:s_2}, see sec.~\ref{s.mixing}) agrees best with the
    analytic solution $u_{\rm e}/u_{\rm gas}=0.379$.}
  \label{f.shock}
 \end{figure}

\subsubsection{Synchrotron}

In Gaussian-cgs units the energy mean emission opacity is
\be
\kappa_{\rm P, e}^{\rm (sy)} \rho =  1.59 \times 10^{-30} n_{\rm e} B^{2} T_{\rm e}^{-2} \ {\rm [ cm^{-1} ] }  ,
\ee
where $B^2 = 4 \pi b^\mu b_\mu$. Numerical fits for the synchrotron opacities are given in terms of the dimensionless parameter $\zeta$,
\be
\zeta = \frac{4 \pi m_{\rm e}^3 c^5}{3 e k h} \frac{T_{\rm r}}{B T_{\rm e}^2}  .
\ee
The energy mean absorption opacity is numerically fitted as
\be
\kappa_{\rm P, a}^{\rm (sy)} = \kappa_{\rm P, e}^{\rm (sy)} \xi^{-3} \left(1 + 5.444 \zeta^{-2/3} + 7.218 \zeta^{-4/3} \right)^{-1} ,
\ee
while the Rosseland mean opacity is given as
\be
\kappa_{\rm R}^{\rm (sy)} = \kappa_{\rm P, e}^{\rm (sy)} \xi^{-3} \cdot  3.24 \times 10^{-2} \zeta^{1.31} \exp \left(-1.60\zeta^{0.463} \right) .
\ee
We additionally fit the number mean opacity which is used in the photon density evolution equation,
\be
\kappa_{\rm n}^{\rm (sy)} = \kappa_{\rm P, e}^{\rm (sy)} \xi^{-3} \cdot 0.868 \zeta \left(1 +0.589 \zeta^{-1/3} + 0.087 \zeta^{-2/3} \right)^{-1}.
\label{eq:numopac}
\ee
Corresponding opacities are derived in the Appendix \ref{ap.opacities}, Eqs. \ref{eq:CalcSyEmi}-\ref{eq:CalcSyNum}.

\subsubsection{Scattering and Comptonization}
The Comptonization term $\widehat{G^0}_{\rm Comp}$ is given by
\cite{sadowski+compt}, and is a function of $T_{\rm e}$ and $T_{\rm
  r}$, scaled by the scattering opacity $\kappa_{\rm es}$. For
$\kappa_{\rm es}$, we use the following energy and angle averaged
Klein-Nishina formula \citep{buchler76, paczynski83}, for consistency,
\begin{equation}
\kappa_{\rm es} = \kappa_{\rm T} \left[ 1 + \left(\frac{T_{\rm r}}{4.5 \times 10^8 K} \right)\right]^{0.86} \ {\rm [ cm^2 g^{-1}]} .
\label{eq:KN}
\end{equation}
In practice, for the radiation temperatures $T_{\rm r}$ that are found in the presented simulations, the Klein-Nishina correction to the Thomson scattering, $\kappa_{\rm T} = 0.2(1 + X) $, is unimportant.

\subsection{Photon number density evolution}
\label{subs:photDens}
We follow the photon number density evolution scheme proposed by
\cite{sadowski+compt}, but using the above improved model of
opacities. Hence, the number density of photons is evolved with the following equation
\be 
\widehat{\dot{n}} = \widehat{\dot{n}}_{\rm sy} + \widehat{\dot{n}}_{\rm ff}  - \left(\kappa^{\rm (sy)}_{\rm n} + \kappa^{\rm (ff)}_{\rm n} \right) \rho \widehat{n} c ,
\label{eq:photondens}
\ee
representing the balance between emission and absorption of
photons. The synchrotron emission of photons is evaluated as
\be
\widehat{\dot{n}}_{\rm sy} = 1.44 \times 10^{5} B n_{\rm e} \ {\rm [cm^{-3} s^{-1}]},
\ee
and we use the following approximation for the bremsstrahlung process
\be
\widehat{\dot{n}}_{\rm ff} = \frac{\kappa^{\rm (ff)}_{\rm P,e} \, \rho  4 \pi \widehat{B}}{2.7012k T_{\rm e}} .
\ee
The absorption term in Eq. \ref{eq:photondens} consists of a~properly integrated and numerically fitted number mean opacity for the synchrotron process, Eq. \ref{eq:numopac}, and bremsstrahlung number mean opacity, approximated with the absorption energy mean opacity given in Eq. \ref{eq:OpacEnergyAbsFF}, $ \kappa^{\rm (ff)}_{\rm n} = \kappa^{\rm (ff)}_{\rm P, a}$. See Appendix \ref{ap.opacities} for details.

\section{Numerical methods}

\label{s.method}

We use the GRRMHD code \texttt{KORAL} \citep{sadowski+koral,sadowski+koral2}.  In
the original version of the code,
gas, magnetic field, and radiation are evolved in a fixed
spacetime described by an arbitrary metric. The magnetic field is
evolved under the assumption of ideal MHD, i.e., assuming that the
electric field vanishes in the gas comoving frame, and without any explicit
resistive term. The radiation field is 
described through the
radiative energy density, its momentum vector (radiation flux) and the photon number
density \citep{sadowski+compt}. The radiation
stress-energy tensor is closed using the M1 closure scheme
\citep{levermore84, sadowski+koral} modified according to
\cite{sadowski+dynamo} to include 
radiative viscosity.

In this work we improve the physics implemented in \texttt{KORAL} and
evolve the electron and ion species independently. 
Below we discuss some implementation details.

\subsection{Adiabatic evolution and viscous dissipation}
\label{s.mixing}

To estimate the energy dissipation in a given cell during a given
  time step, we compare the thermal energy in the cell, as obtained
  via the energy equation, with the thermal energy that is predicted
  by {\it purely adiabatic evolution}. The difference between these
  two quantities reflects the non-adiabatic increase in the internal
  energy of the gas, i.e, the ``viscous'' dissipation.

For computing the adiabatic evolution, we use the separate entropy
evolution equations of the electrons and ions, viz., (\ref{e.entr1})
and (\ref{e.entr2}), with the right-hand sides set to zero. However,
there is a somewhat subtle issue here. Consider the finite-difference
version of the adiabatic evolution equation of one of the species in
cell $i$ between time $n$ and $n+1$:
\be (\rho s u^t)^{n+1}_{i}=(\rho s u^t)^{n}_{i} - \frac{\Delta
  t}{\Delta x}\left((\rho s u^x)^{n}_{i+1/2}-(\rho s
u^x)^{n}_{i-1/2}\right). \label{eq:s_simple} \ee Here, $s$ is the
entropy per particle of the species under consideration, and the
fluxes are computed at the cell walls $i\pm1/2$ in the usual way. The
above equation states that the entropy content of the cell at time
$n+1$ is a linear combination of the old entropy in the cell and new
entropy that has flowed into (or out of) the cell. This can equally be
written as
\begin{eqnarray}
s^{n+1}_{i} &=& \frac{(\rho u^t)^{n}_{i}}{(\rho u^t)^{n+1}_{i}}
s^{n}_{i}- \frac{\frac{\Delta t}{\Delta x}(\rho
  u^x)^{n}_{i+1/2}}{(\rho u^t)^{n+1}_{i}} s^{n}_{i+1/2}+
\frac{\frac{\Delta t}{\Delta x}(\rho u^x)^{n}_{i-1/2}}{(\rho
  u^t)^{n+1}_{i}} s^{n}_{i-1/2}. \label{eq:s_1}\\ &\equiv& f_i
s^{n}_{i}+ f_{i+1/2}
s^{n}_{i+1/2}+f_{i-1/2}s^{n}_{i-1/2}, \label{eq:s_2}
\end{eqnarray}
where the three $f$'s are the fractions of the final state represented
by the three contributing parcels of gas.

Equation (\ref{eq:s_2}) is incorrect when we consider finite-sized
  cells and finite time steps. Whenever two parcels of gas with
  different properties
are mixed together at
constant volume, the resulting combined gas will end up 
in a state
whose energy content is given by the sum of the initial energies of
the two parcels (energy is conserved), but whose entropy will not be
the sum of the entropies of the two initial parcels (entropy is not
conserved). In effect, thermal energy will flow from particles of the
hotter parcel to those of the cooler parcel, conserving thermal
energy, but causing the net entropy to increase.

The relevant question now is the following: Does the entropy
  increase caused by the transfer of thermal energy between two or
  more parcels of gas inside a single cell represent viscous
  dissipation?  We think not. This entropy increase is caused by the
  irreversible process of transfering energy from hot to cold gas
  within a cell. It takes place merely because our numerical code is
  unable to keep track of the properties of individual parcels within
  the cell but is forced to consider a single homogenized fluid. There
  is no real dissipation involved, i.e., no creation of new thermal
  energy via shocks or turbulence or magnetic reconnection. In other
  words: The finite-volume cell-averaging process artificially
  increases the entropy of a zone.  This increase of entropy should
  not be treated as dissipation.

We further argue that it is the above thermally homogenized fluid,
  which has enhanced entropy due to thermal mixing, that represents
  what we call \textit{ purely adiabatic evolution} in the first
  paragraph of this subsection. It is this thermal energy that should
  be subtracted from the thermal energy obtained via the energy
  equation to estimate the energy dissipation in the cell.  The
  resulting estimate of the dissipation will be less than the amount
  we would estimate if we directly took $s_i^{n+1}$ from
  equation~(\ref{eq:s_2}) and computed an internal energy from that.

We note yet another complication. Consider for simplicity a single species
  ideal gas with a constant adiabatic index $\gamma$. The entropy per
  particle of the gas is proportional to $s_1 = \ln(p/\rho^\gamma)$,
  but sometimes, the quantity $s_2 = p/\rho^{\gamma-1}$ itself is
  taken as the conserved quantity
  \citep[e.g.,][]{ressler+15}. Let us imagine that we ignore the
  entropy of thermal mixing discussed above and estimate the internal
  energy corresponding to \textit{purely adiabatic evolution} directly
  from the entropy $s^{n+1}$ in equation~(\ref{eq:s_2}). The answer we
  get will be different depending on whether we use $s_1$ or
  $s_2$. The reason is that equation~(\ref{eq:s_2}) is a linear
  combination of three entropy terms. However, each entropy term is a
  non-linear function of the internal energy density $u$ and mass
  density $\rho$ of the particular gas parcel, and the function is
  different in the case of $s_1$ and $s_2$. As a result, the final
  estimate of $u$ for the homogenized cell will be different whether
  we use $s_1$ or $s_2$.

The solution to all the above complications is not obvious. Recall
  that the problem is caused by the fact that, in any numerical
  scheme, we can keep track of only a limited set of data in each
  cell, viz., the gas properties at cell centers. Once a piece of gas
  has crossed into a cell, it loses its individuality and is
  considered to be mixed homogeneously with the rest. Exactly how
  should we mix the parcels when defining what we call \textit{purely
    adiabatic evolution}? We discuss below in \S\ref{sec:mixing} one
  possible approach that is arguably consistent. We have not used that
  approach in the work reported in this paper. Here we make the
  somewhat arbitrary choice of mixing the different parcels at a
  constant density equal to the final density of the given cell.  That
  is, for each of the three parcels of gas in equation (\ref{eq:s_2}),
  we calculate the internal energy density corresponding to the
  \textit{final} rest-mass density of the cell: $\tilde u_{i, i\pm
    1/2}^{n+1}=u(s^n_{i, i\pm 1/2},\rho^{n+1}_{i})$.  We then linearly
  sum these final internal energies, using the same mixing fractions
  $f$ as before, \be u^{n+1}_{i}=f_i \tilde u^{n+1}_{i}+ f_{i+1/2}
  \tilde u^{n+1}_{i+1/2}+f_{i-1/2}\tilde
  u^{n+1}_{i-1/2}. \label{eq:u_f} \ee This is the internal energy that
  we use as our proxy for ``purely adiabatic evolution.'' Once we have
  $u^{n+1}_i$, we then recover the corresponding temperature,
  $\theta^{n+1}_i=\theta(u^{n+1}_{i},\rho^{n+1}_{i})$, the entropy per
  particle, $s^{n+1}_i=s(u^{n+1}_{i},\rho^{n+1}_{i})$, etc., using the
  relations discussed in Appendix A. Note that the above calculations
  are done for each of the two species.

One final detail concerns the entropies and internal energies,
  $s^n_{i\pm 1/2}$, $\tilde u^n_{i\pm1/2}$, at the cell boundaries.  The
  natural choice would appear to be \be s^{n}_{i\pm1/2}=\frac{(\rho s
    u^x)^{n}_{i\pm1/2}}{(\rho u^x)^{n}_{i\pm1/2}}.  \ee However, we
  have found that even small errors in the numerical reconstructions
  at the walls lead, because of the exponential dependence of energy
  density on entropy, to erroneous values of the internal energy
  density. This in turn often leads to negative temperatures in the
  final state. To avoid this, we use \textit{upwind} values of the
  entropy per particle, \be s^{n}_{i\pm1/2}=s^n_{\rm upwind}.  \ee For
  example, if gas flows from cell $i-1$ to $i$, we set $s^n_{i-1/2} =
  s^n_{i-1}$, while if gas flows from cell $i$ to $i-1$, we set
  $s^n_{i-1/2} = s^n_i$. Similarly for $s^n_{i+1/2}$.  This approach
  is guaranteed to be stable and, as we discuss below, provides good
  accuracy.

\subsubsection{A potentially more consistent approach}\label{sec:mixing}

An unsatisfactory aspect of the method described above is our decision
to mix the individual fluid parcels at constant density. This is
computationally very convenient and is the reason we chose
it. However, one suspects that there ought to be a physically more
consistent approach.

One attractive option is to mix the parcels in such a fashion that
  the internal energy of the final state is a minimum. Let us suppose
  that we have N parcels of fluid, each with a mass $M_j$ and
  entropy $S_j$. We wish to choose the volumes $V_j$ occupied by the
  parcels, subject to the condition that they sum up to the total
  volume $V$ of the cell,
\begin{equation}
\sum_{j=1}^N V_j = V. \label{eq:volume}
\end{equation}
Let the internal energy of each parcel be $U_j$, which is a function
of $V_j$. We are interested in adjusting the $V_j$ so as to minimize
the total internal energy,
\begin{equation}
\sum_{j=1}^N U_j \equiv U = {\rm minimum}.
\end{equation}
Using a Lagrange multiplier $\Lambda$ for the volume constraint,
the problem reduces to minimizing
\begin{equation}
\sum_{j=1}^N \left(U_j + \lambda V_j\right) = {\rm minimum}.
\end{equation}
Each parcel must thus satisfy the condition
\begin{equation}
\left(\frac{\partial U_j}{\partial V_j}\right)_{S_j} = -\lambda,
\end{equation}
where we have explicitly noted that each parcel conserves its own
entropy (it changes adiabatically) under the volume changes we are
considering.  From elementary thermodynamics, $(\partial U/\partial
V)_S = -P$, where $P$ is the pressure. Thus, the minimum internal
energy we seek satisfies
\begin{equation}
P_j = \lambda,
\end{equation}
i.e., all the fluid parcels have the same pressure (which is equal to
$\lambda$). The actual value of this pressure is obtained by imposing
the volume constraint (\ref{eq:volume}).

While the above ``uniform pressure'' result is simple and
  physically appealing, it is not numerically as convenient as the
  constant density prescription we have used in the present paper. The
  reason is that the inversion from pressure to volume for a fixed
  entropy involves a non-linear equation in the quasi-relativistic
  regime (Appendix A), not to mention that each fluid parcel itself
  consists of two fluids, ions and electrons, each with its own
  conserved entropy. Thus, one needs to solve a set of coupled
  nonlinear equations to compute the constant-pressure minimum-energy
  state. We intend to explore this in future work.

\subsection{Other implementation notes}
\label{s.implementation}

Within each Runge-Kutta time substep the evolution equations are
applied via the following sequence of operations.

(i) The effective adiabatic index of the gas, $\gamma_{\rm gas}$, is
calculated through Eq.~\ref{e.gammagas} using the electron and ion
temperatures from the end of the previous time step.

(ii) The advective parts (no source terms) of
Eqs.~\ref{eq.cons1}-\ref{eq.cons4} (total gas) are applied in the
standard explicit way (for details of the particular implementation
see \citealp{sadowski+koral} ). In the case of
Eqs.~\ref{e.entr1}-\ref{e.entr2} (electron and ion entropy), we follow
the procedure described in the previous sub-section.

(iii) At the end of the explicit operator we calculate at each cell the
local viscous dissipation rate by comparing the non-adiabatic
evolution of the total gas with the adiabatic evolution of electrons
and ions using Eq.~\ref{e.qv}.

(iv) Knowing the electron and ion temperatures, and the magnetic to
gas pressure ratio, the fraction of the viscous heating applied to the
electrons is calculated according to Eq.~\ref{e.deltae}. 

(v) Electron and ion entropies are updated by increasing their energy
densities by the corresponding fraction of the energy injected into the
gas by the viscous dissipation. 

(vi) The effective adiabatic index of the gas, $\gamma_{\rm gas}$, is recomputed using the just obtained species temperatures.

(vii) Steps (iv) to (vi) are iterated three times to allow
$\gamma_{\rm gas}$ to converge to the correct value corresponding to
the final, heated state of electrons and ions.

(viii) The remaining source terms in
Eqs.~\ref{eq.cons1}-\ref{eq.cons4} and
Eqs.~\ref{e.entr1}-\ref{e.entr2}, i.e., the radiative coupling,
$G_{\rm \nu}$, the photon source term, $\dot n$, and the Coulomb
coupling, $q^{\rm C}$ are applied through a semi-implicit operator in
a way similar to the one described in \citep{sadowski+koral2}. Six
primitive quantities are iterated to find the final state: energy
density and three components of velocity (corresponding to either gas
or radiation), photon number density and electron entropy density. At
each iteration, the gas density and ion temperature are obtained from
the conserved quantity, $\rho u^t$, and energy conservation ($u_{\rm
  e}+u_{\rm i}=u_{\rm gas}$), respectively.

For the sake of stability, regular limits are applied on the evolved
quantities. In particular, neither the electron nor ion temperature is
allowed to drop below one percent of the gas temperature (or, equally, for
  the pure hydrogen, that neither temperature can exceed $1.99$ times
  the gas temperature). We find
that this criterion is very rarely met in the simulations and that this particular choice of the limit does not influence
the accretion flow structure. When imposing density floors (which rarely happened for
the simulations presented here), we kept the
internal energy of electrons and ions
fixed to avoid adding any artificial heating.

\section{Tests}
\label{s.tests}

In this Section we describe two sets of test problems which verify the
implementation of the new physics in the \texttt{KORAL} code, i.e.,
the treatment of viscous heating and its effect on electrons and ions,
and the radiative and Coulomb coupling between electrons, ions, and radiation.

\subsection{1D Shock test}
\label{s.shock}

We start with a one-dimensional high Mach number shock test problem as
described in detail in \cite{ressler+15}. Two fluids move in opposite
directions with $|v|=10^{-3}c$ and collide at $x=0$.  The gas
adiabatic index is fixed to $\gamma=5/3$, while the electron adiabatic
index is $\gamma_{\rm e}=4/3$ (\citealt{ressler+15} also considered
the trivial case $\gamma_{\rm e}=5/3$, but we focus here on the more
difficult problem). The temperature of the gas is set up to give a
Mach number of ${\cal M}=49$. We identify dissipation by following the
entropy of the gas \citep[as in][]{ressler+15}, not the entropies of
the species (as we propose in this paper). We choose a resolution of 512
grid cells.

Figure~\ref{f.shock} presents the electron to ion energy density ratio
profile at a time corresponding to the shock having travelled a
distance $\Delta x\approx 0.13$ in arbitrary units. The central part
of the plot shows the properties of the shocked fluid. As
\cite{ressler+15} have shown, one should expect $u_{\rm e}/u_{\rm
  gas}=0.379$. Adopting a linear formula for the entropy per
particle\footnote{All the formulae adopted for this test are valid
  only for constant adiabatic indices, in contrast to the more general
  entropy prescription introduced in this paper.} ($s=p/\rho^\gamma$),
we obtain a post-shock $u_{\rm e}/u_{\rm gas}=0.406$, within $7\%$ of
the correct solution. Using the logarithmic formula
($s=\log{p/\rho^\gamma}$) and evolving it using equation
(\ref{eq:s_simple}) blindly, we obtain $u_{\rm e}/u_{\rm gas}=0.336$
($11\%$ error). Finally, adopting the logarithmic prescription, but
evolving it with the modified scheme described in equation
(\ref{eq:s_2}, sec.~\ref{s.mixing}), we get $u_{\rm e}/u_{\rm
  gas}=0.397$ ($5\%$ error), which is closest to the correct solution.

\subsection{Driven turbulence}
\label{s.driven}

\begin{figure}
 \includegraphics[width=1.\columnwidth]{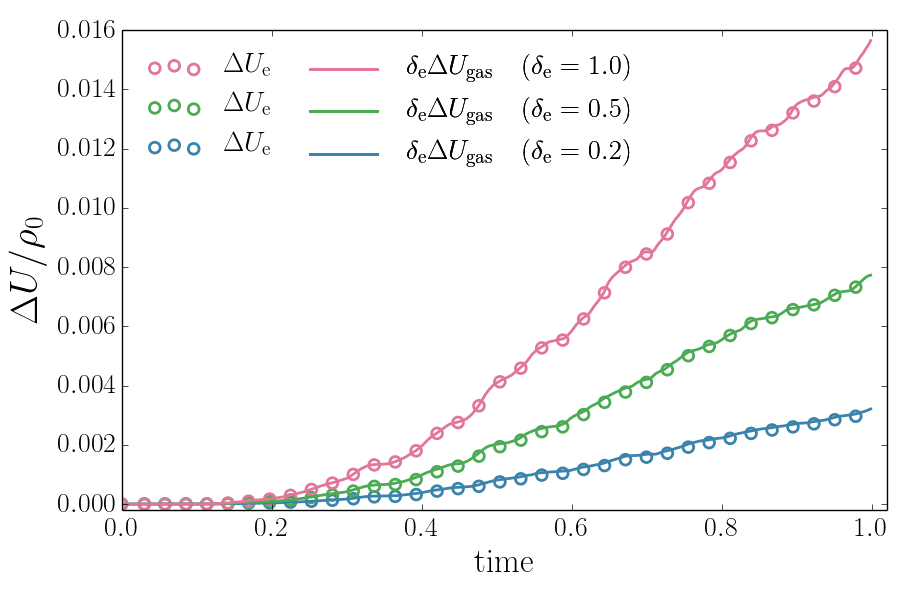}\vspace{-.85cm}
 \includegraphics[width=1.\columnwidth]{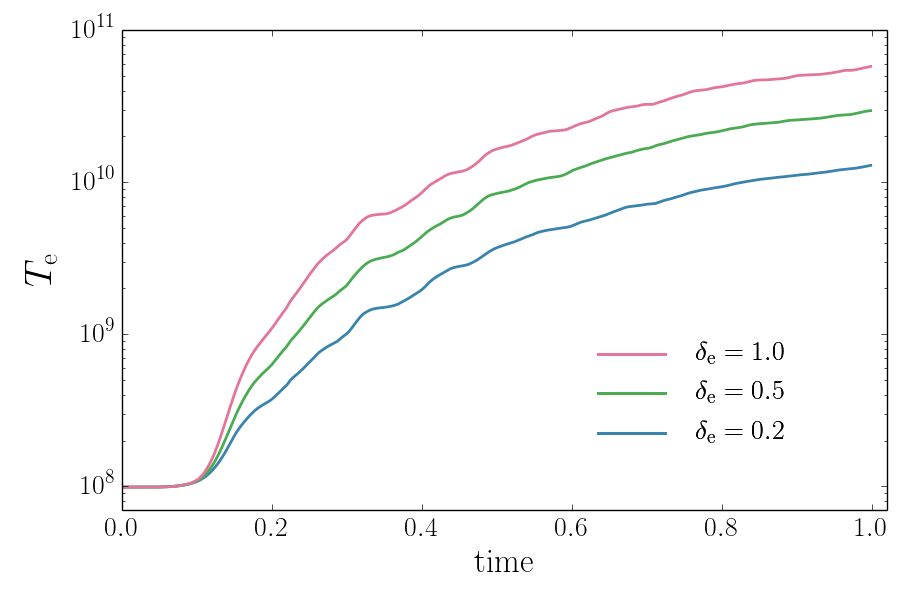}\vspace{-.85cm}
   \caption{Top panel: Increase of the energy content in electrons in the driven
     turbulence test (Section~\ref{s.driven}). The colors correspond
     to three independent runs with different fractions of viscous
     heating going to electrons: pink, green and blue correspond to
     $\delta_{\rm e}=1.0$, $\delta_{\rm e}=0.5$, and $\delta_{\rm e}=0.2$,
     respectively. Circles denote the actual increase of energy of
     electrons. The solid lines reflect the corresponding fraction of
     the total gas energy increase. The fact that energy in electrons
     traces the given fraction of the total energy proves that the viscous
     heating is applied properly to the different species. Bottom panel:
The corresponding increase in the electron temperature. In all cases electrons
 transition from sub- to relativistic regime.
Time is given in arbitrary units.}
\label{f.driven}
\end{figure}

To validate our implementation of viscous heating of electrons and ions, we performed a test similar in spirit to the
MHD driven turbulence test of \cite{ressler+15}\footnote{Our version uses consistent adiabatic indices of gas, ions, and
  electrons instead of fixing them at predetermined values. We also use
a slightly  different perturbation magnitude.} 
We set up a uniform
medium on a two dimensional plane with gas density, $\rho_{\rm 0}$,
species
temperature, $T_{\rm e,0}=T_{\rm i,0}=10^{8}\rm\, K$, horizontal magnetic
field with gas to magnetic pressure ratio $\beta=p_{\rm gas}/p_{\rm mag}=6$, and
initially zero velocities. At every timestep we perturbed the velocities
with random, divergenceless, Gaussian velocity perturbations with average power
spectrum, $P(|\delta v|^2)\propto k^6 \exp(-8k/k_{\rm peak})$ with
characteristic wavevector, $k_{\rm peak}$, corresponding to half
of the computational box, $k_{\rm peak}=4\pi/L$. The normalization was
chosen to provide a saturated rms velocity $v_{\rm rms}\approx 0.08c$.
For constructing the
perturbation velocity field we followed \cite{dubinski+95}. We also
made sure the perturbations did not introduce net momentum by
substracting uniformly from all cells the non-zero residual.

We performed three independent simulations, each on a two-dimensional
grid of $128 \times 128$ cells. 
The random velocity perturbations imposed at
every time step introduced kinetic energy into the system which
dissipates and contributes finally to the gas thermal energy. As a result, the
gas energy content increases with time.
Once saturation is
reached (i.e., the dissapation rate matches the energy injection rate), the increase is linear.

A fraction, $\delta_{\rm e}$, of this non-adiabatic heating
is deposited in
electrons according to Eq.~\ref{e.entr1}. The remaining part goes into
ions (Eq.~\ref{e.entr2}). In the top panel of Fig.~\ref{f.driven} we show with open
circles the increase
of the domain-integrated (total) internal energy of electrons ($\Delta U_{\rm e}$)
 in the simulations assuming $\delta_{\rm e}=1.0$
(pink markers), $\delta_{\rm e}=0.5$ (green), and $\delta_{\rm e}=0.2$ (blue). This increase is compared with the corresponding fraction of
the increase of gas thermal energy ($\delta_{\rm e}\Delta U_{\rm gas}$), 
denoted by solid
lines. The fact that the internal energy of electrons increases in all
three cases at the same rate as the corresponding fraction of the
total energy proves that the non-adiabatic viscous heating is
identified and applied properly in the code. The bottom panel shows the corresponding
increase of the electron temperature. Electrons, initially at $10^8\rm K$, heat up and reach
$\sim 10^{10}\rm K$ near the end of the test simulations, transitioning from the 
sub-relativistic to relativistic regime.

\subsection{Radiative and Coulomb coupling of species}
\label{s.box}

We consider an isolated system, a box filled with ions, electrons, and
radiation. Assuming that all physical quantities are spatially
constant, the only relevant equations are the energy components of
Eqs. \ref{eq.cons2}-\ref{eq.cons3}, the electron entropy evolution
(Eq. \ref{e.entr1}), and the photon density evolution
(Eq. \ref{eq.cons4}). In the absence of dissipative heating these four
formulae can be simplified to the~following system of ordinary
differential equations for the relevant temperatures,
\begin{align}
\label{eq:SimpSys1}
& \dot{T}_{\rm e} = \frac{m_p (\gamma-1)} {k \rho} \left( \widehat G^t + q^C \right) , \\
&  \dot{T}_{\rm i} = -\frac{m_p (\gamma-1)}{k \rho} q^C , \\
& \dot{T}_{\rm r} = \frac{\epsilon T_{\rm r}}{\widehat{E} }  \left(1 - \frac{T_{\rm r}}{T_{\rm e}} \right) , \\
& \dot{T}_n = -\frac{T_n \rho}{3 \widehat{E}} \left(\kappa_{\rm P, a} \, \widehat E c - \frac{T_{\rm r}}{T_{\rm e}} \epsilon \right) ,
\label{eq:SimpSys4}
\end{align}
where we explicitly assumed for simplicity that all the adiabatic
indices are fixed and equal $\gamma$. The energy coupling term,
$\widehat G^t$, equals,
\be 
\widehat G^t = \rho \kappa_{\rm P, a} \, \widehat E c - \epsilon  = \rho \left(  \kappa_{\rm P, a} \, 2.7012 k c  n T_{\rm r}  -4 \sigma \kappa_{\rm P, e} T_{\rm e}^4 \, \right) ,
\ee
\begin{table}
\begin{center}
\caption{Parameters of the "gas and radiation box" simulations}
\label{tab:ode}
\begin{tabular}{lllllll}
\hline
\hline
Name &
$T_{\rm e} {\rm [K]}$ & $T_{\rm i} {\rm [K]}$ &
$T_{\rm r} {\rm [K]}$ & $T_n {\rm [K]}$ & $\rho \, {\rm [ g \, cm^{-3} ]}$\\
\hline
\hline
\texttt{ODE1} & $10^{10}$ & $10^{11}$ & $10^{9}$ & $10^{9}$ &$10^{-4}$\\
\texttt{ODE2} & $3\times 10^{9}$ & $10^{10}$ & $10^{9}$ & $5 \times 10^{8}$ & $1$\\
\texttt{ODE3} & $10^{8}$ & $10^8$& $ 10^{9}$  & $10^{9}$ & $10^{-10}$ \\
\hline
\hline
\end{tabular}
\end{center}
\end{table}
and the temperature $T_n$ is related to the number density of photons,
\be 
T_n =  \left( \frac{2.7012 c k}{4 \sigma}  n \right)^{1/3} \le T_{\rm r},
\ee
and represents the black body temperature corresponding to a given
photon density $n$ (if the radiation has a~black body spectral energy
distribution, then simply $T_n = T_{\rm r}$). We set
$\gamma = 5/3$ and account only for bremsstrahlung radiation and
Coulomb coupling. 

\begin{figure}
 \includegraphics[width=1.\columnwidth]{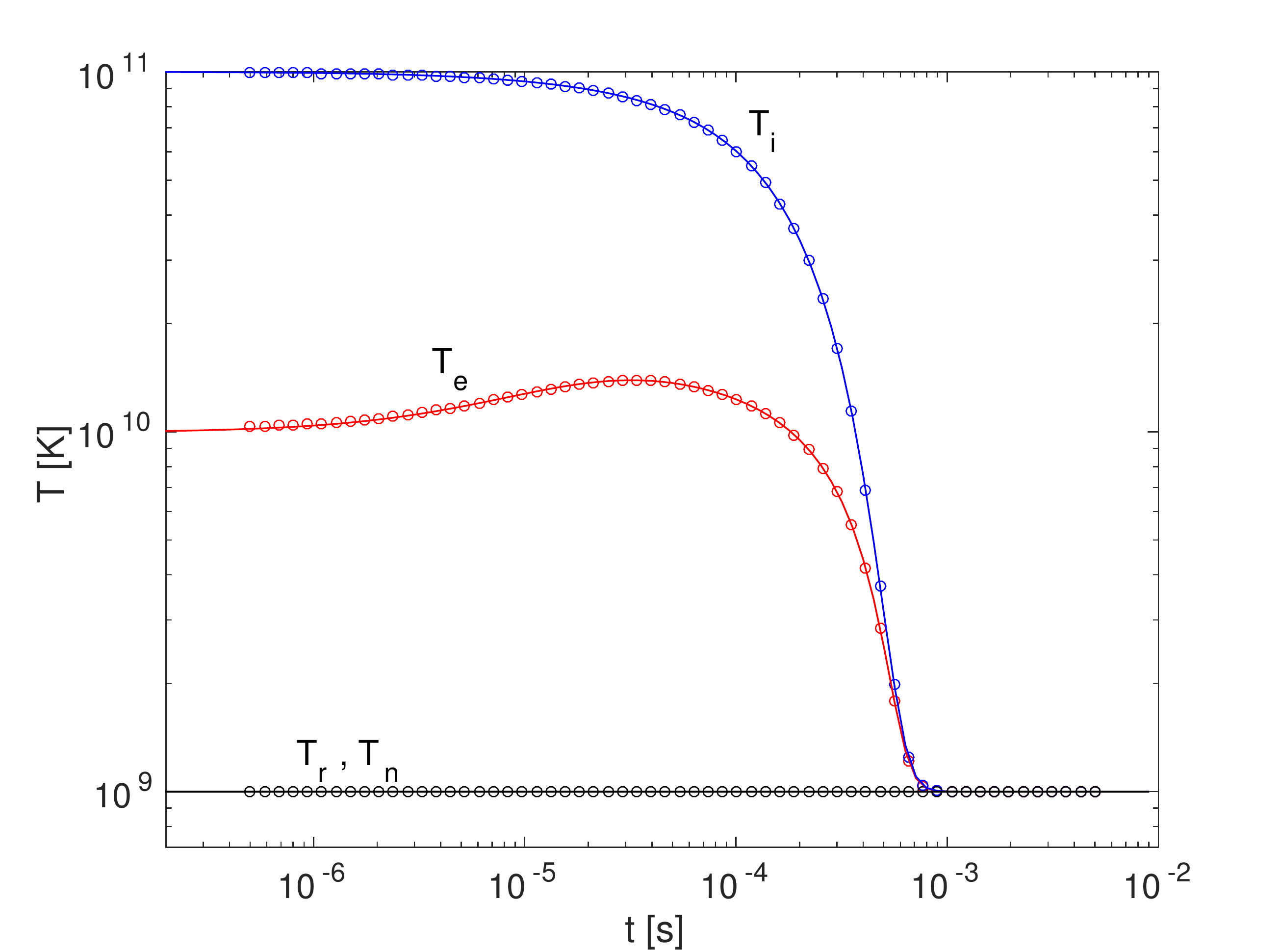}
  \includegraphics[width=1.\columnwidth]{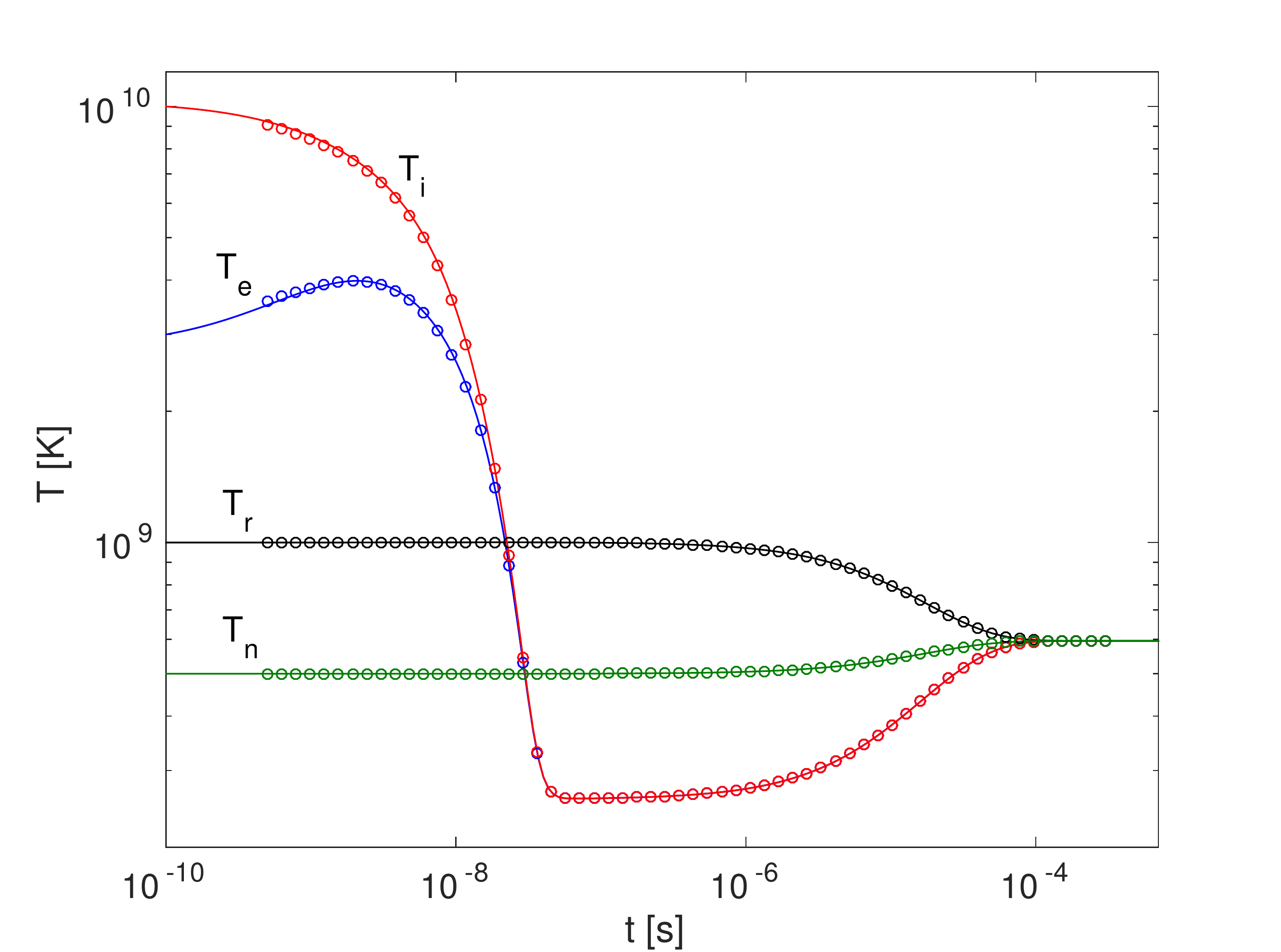}
   \includegraphics[width=1.\columnwidth]{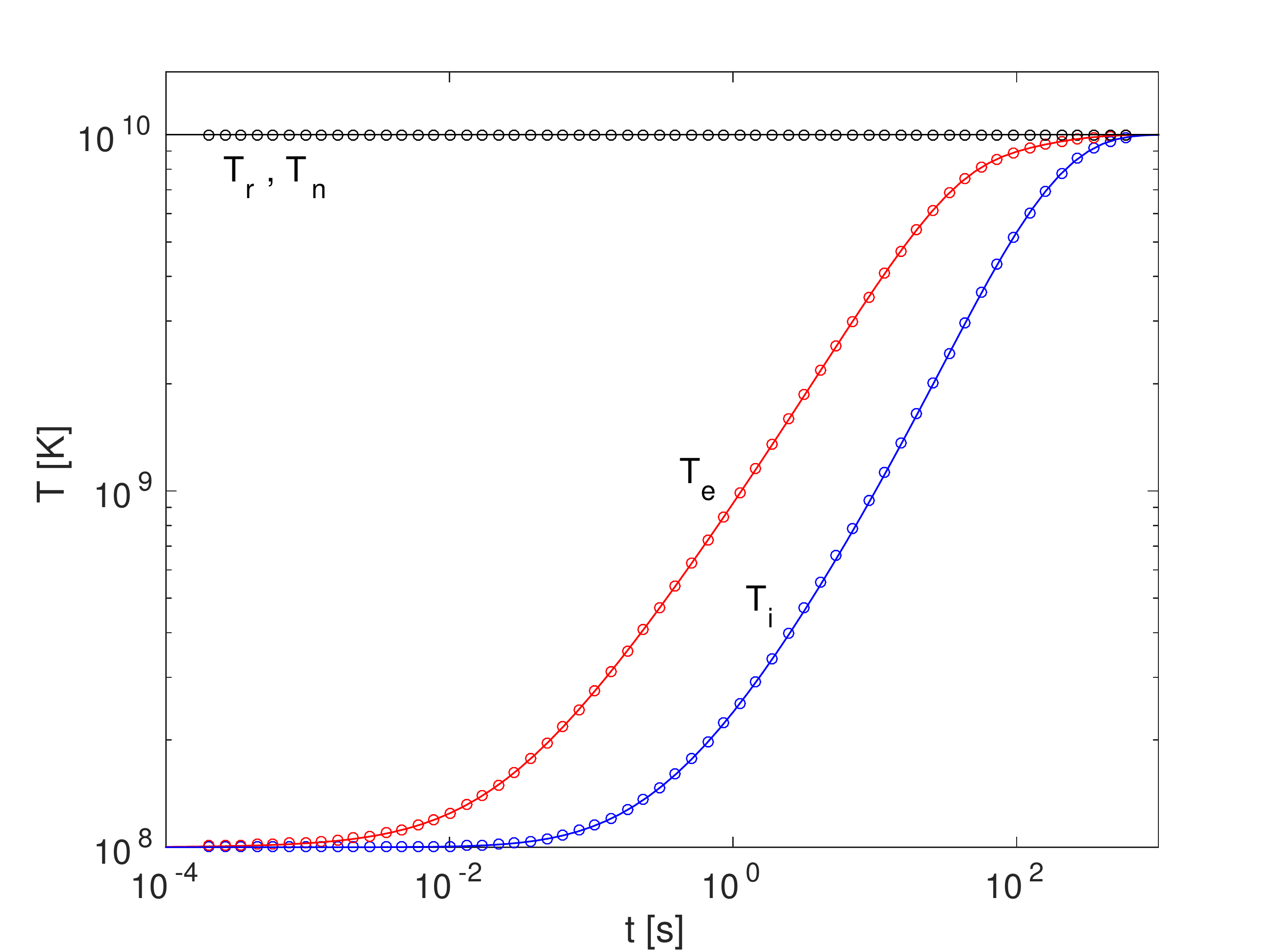}
   \caption{Evolution of the temperatures $T_{\rm e}$, $T_{\rm i}$,
     $T_{\rm r}$, and $T_{\rm n}$ in the "gas and radiation box" test
     simulations (Section~\ref{s.box}). Continuous lines represent
     results of the Runge-Kutta solver applied directly to
     Eqs.~\ref{eq:SimpSys1}-\ref{eq:SimpSys4}, dots represent
     solutions found with \texttt{KORAL} via its semi-implicit
     solver. In \texttt{ODE1} (top) electrons are radiatively cooled
     and kinetically heated by the hotter ions. In \texttt{ODE2}
     (middle) additionally radiation is imbalanced, with $T_{\rm r} >
     T_{\rm n}$. Because of the large gas density and efficient
     Coulomb coupling, electrons and ions quickly equilibrate to the same
     temperature, only later being brought in balance with the
     radiation. In \texttt{ODE3} (bottom), the
     electrons are heated up by absorption of hotter radiation and the
     collisionally coupled ions follow with a~delay. At late times in
     all three tests, the systems approach thermodynamical equilibrium
   with $T_{\rm e}=T_{\rm i}=T_{\rm r}=T_{\rm n}$.}
 \label{fig:modelsODE}
\end{figure}

We performed three runs, for which the initial conditions are summarized in
Table~\ref{tab:ode}. The evolution of the four temperatures in these simulations is plotted in Fig. \ref{fig:modelsODE}. For
the reference solution we take the accurate results produced by the
implicit Runge-Kutta solver \citep{solverODE}, applied directly to the
system of equations
\ref{eq:SimpSys1}-\ref{eq:SimpSys4}. Fig. \ref{fig:modelsODE}
indicates that the \texttt{KORAL} code, which solves the corresponding equations in
their original form during the semi-implicit operator stage, is able to reproduce the Runge-Kutta solutions with very small error.

\section{Low luminosity accretion flows}
\label{s.low}

\begin{figure}

 \centering\includegraphics[width=1.08\columnwidth]{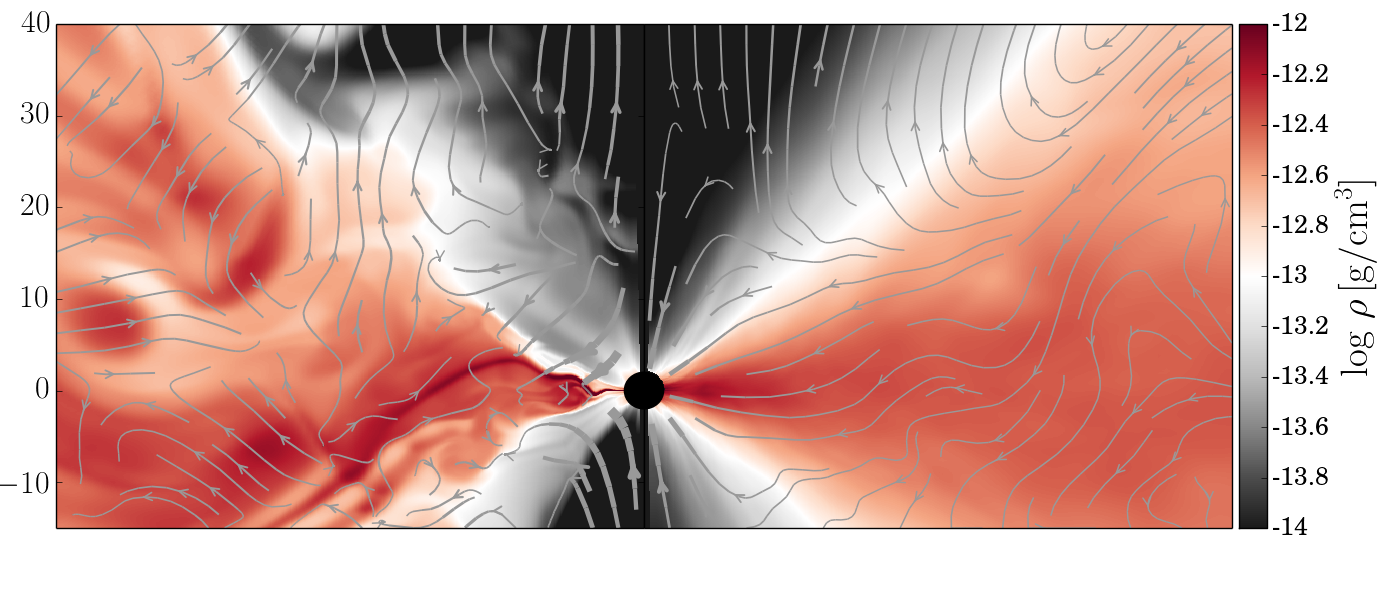}\vspace{-.40cm}
 \includegraphics[width=1.08\columnwidth]{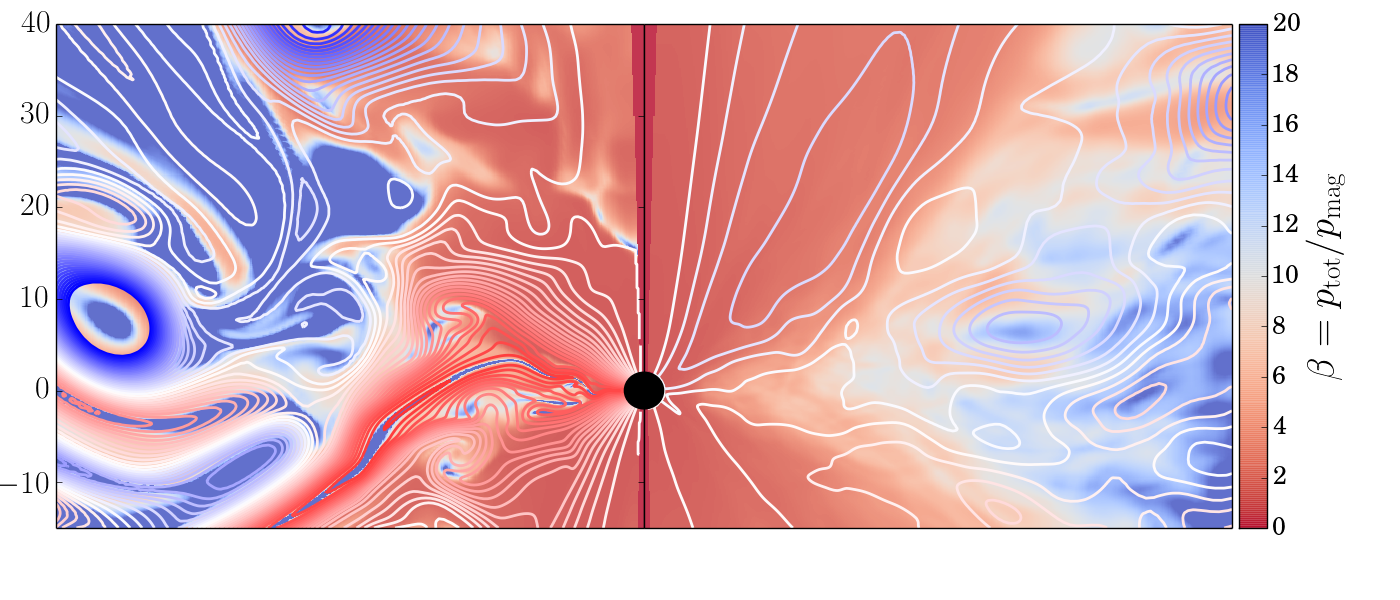}\vspace{-.40cm}
 \includegraphics[width=1.08\columnwidth]{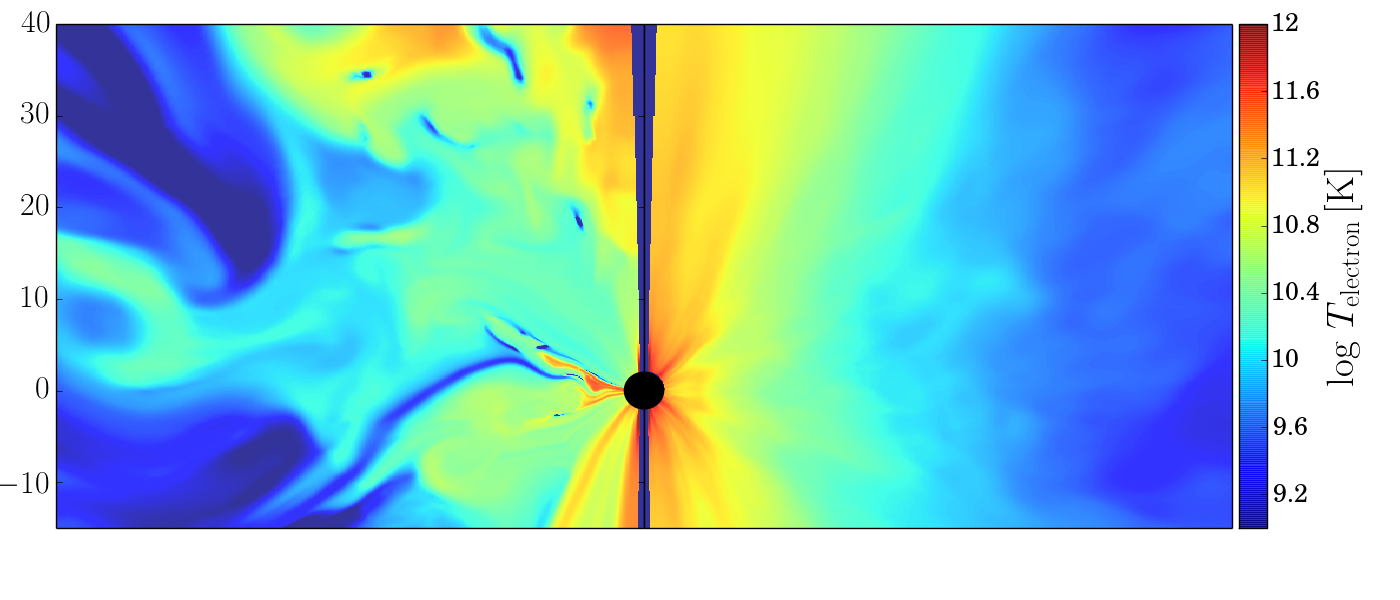}\vspace{-.40cm}
 \includegraphics[width=1.08\columnwidth]{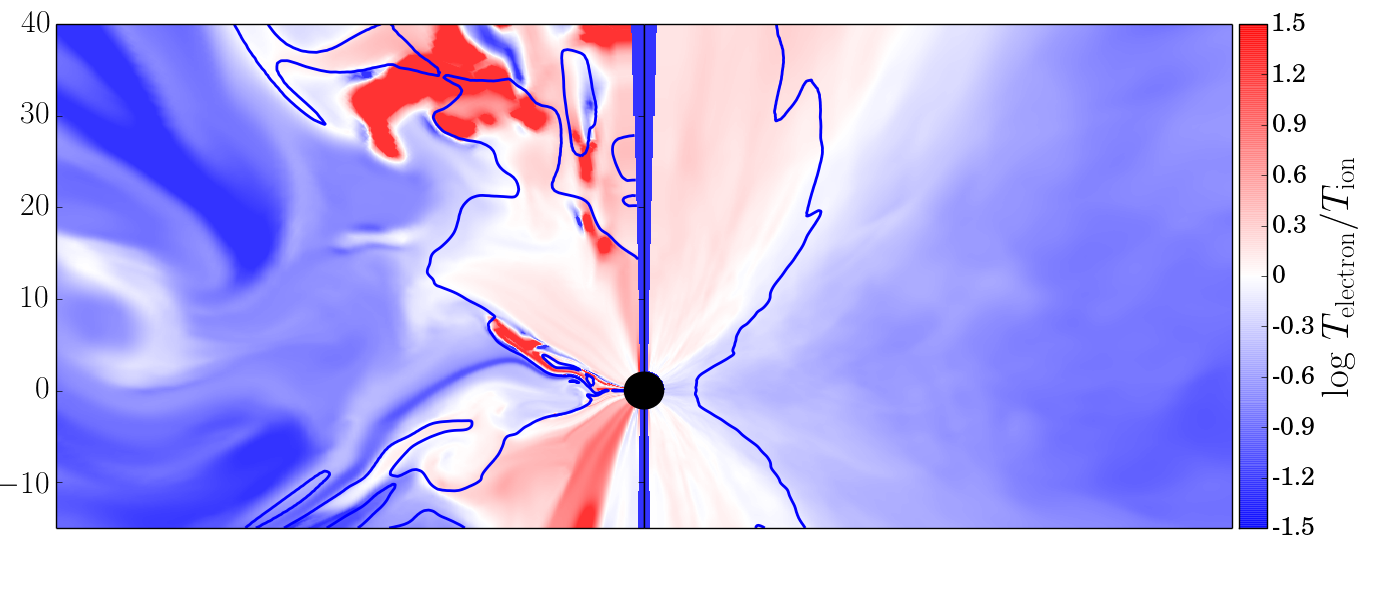}\vspace{-.40cm}
 \includegraphics[width=1.08\columnwidth]{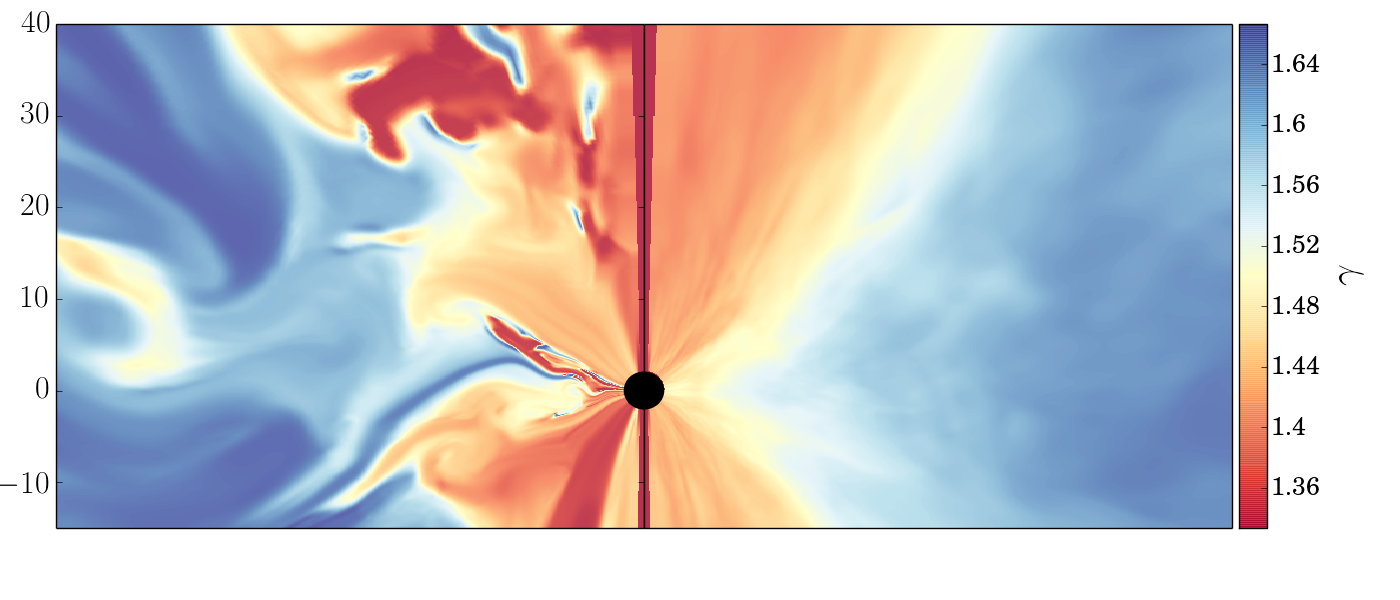}\vspace{-.40cm}
 \includegraphics[width=1.08\columnwidth]{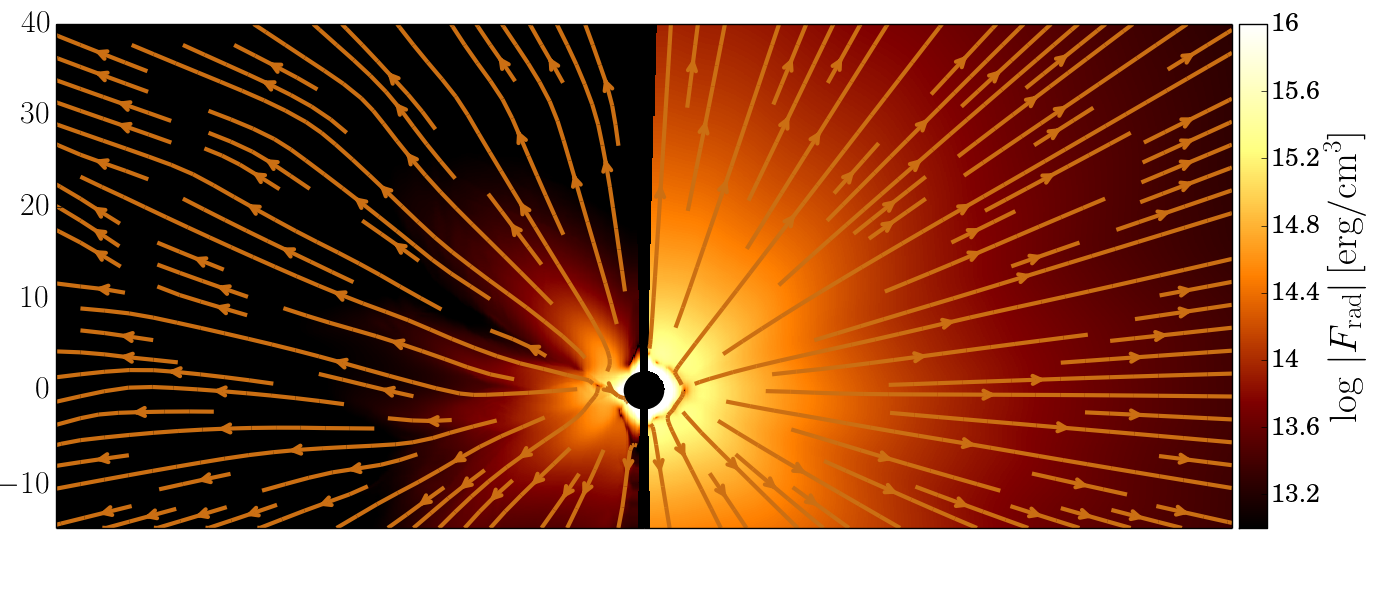}\vspace{-.40cm}
 \includegraphics[width=1.08\columnwidth]{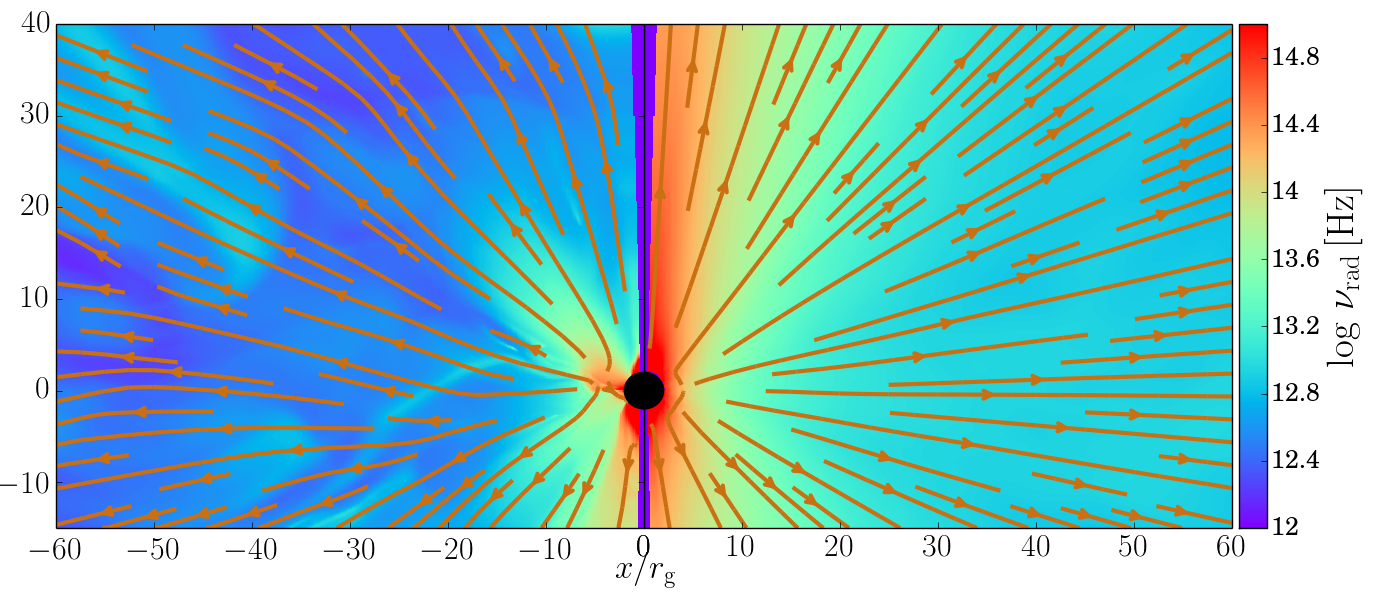}\vspace{-.40cm}
  \caption{Snapshot (left) and time-averaged (right) distributions of
    (top to bottom) gas density, magnetization and poloidal magnetic
    field lines, electron temperature, electron to ion temperature
    ratio, effective gas adiabatic index, magnitude of the radiative
    flux, and the characteristic radiation frequency, for model
    \texttt{Rad8}. Streamlines in the topmost and two bottommost
    panels show the directions of the poloidal velocity and radiative
    flux, respectively. The blue contours in the fourth panel show
    where the magnetic pressure equals the gas pressure.  }
 \label{f.d250}
\end{figure}

In this Section we describe a series of simulations of low luminosity
black hole accretion flows performed with the
methods described above. The new physics includes independent
evolution of electrons and ions with individual viscous heating, Coulomb
coupling and radiative loses, and a sophisticated
treatment of radiation which is coupled to electrons through
spectrally integrated free-free and synchrotron opacities, sensitive
to the characteristic temperature of the local radiation field. We also
distinguish between energy mean (Planck) and Rosseland mean opacity
coefficients. All these elements, together with the GR treatment of MHD
already implemented in \texttt{KORAL}, put us in a position to 
study the accretion flows with intermediate optical depths.


In this paper we discuss a couple of examples of how the BH mass,
radiative cooling, and the treatment of the adiabatic index affect the
properties of the accretion flows. In a follow up study we will investigate
in detail the transition induced by increasing accretion rate.

\subsection{Models}

We have simulated five models, the parameters of which are specified in
Table~\ref{t.models}. In three of these (\ttt{Rad8},
\ttt{Rad8SMBH}, \ttt{Rad4}), the gas was coupled to radiation and experienced
radiative losses and the momentum transfer. In the remaining two models
(\ttt{Hd8}, \ttt{Hd8fg}) only the hydrodynamical quantities were evolved
and radiative effects were neglected. In all the models we
accounted for Coulomb coupling.

We chose $10\msun$ as our fiducial BH mass, but one of the models
(\ttt{Rad8SMBH}) considered a BH of mass $4\times 10^6\msun$, roughly
corresponding to the mass of the supermassive BH in the center of our
Galaxy \citep[e.g.,][]{gillessen+09}.

All but one run simulated accretion flows with an average accretion
rate of a ${\rm few }\times 10^{-8}\dot M_{\rm Edd}$. Simulation
\ttt{Rad4} is the exception corresponding to a much higher accretion
rate of a ${\rm few }\times 10^{-4}\dot M_{\rm Edd}$.

Two of the simulations (\ttt{Hd8} and \ttt{Hd8fg}) are
non-radiative. They differ from each other in the treatment of the
adiabatic index. While the former recalculates the adiabatic index of
the gas at each time step and uses the generalized form of entropy as
introduced in this work (Appendix A), the latter assumes fixed values
of the gas and electron adiabatic indices ($5/3$ and $4/3$,
respectively). Correspondingly, the dissipation is obtained from the
entropy of the total gas, not from the entropies of the individual
species. This corresponds to the method adopted in \cite{ressler+15}.

\subsection{Numerical setup}
\label{s.setup}

Every simulation was performed in the Kerr-Schild metric and assumed a
non-rotating BH. We restricted ourselves to axisymmetry and applied
the fiducial model of the mean-field dynamo from
\cite{sadowski+dynamo} to prevent the axisymmetric magnetic field from
decaying.  We used spherical-like coordinates and adopted a resolution
of 384x384x1 cells in radius, polar angle and azimuth,
respectively. The radial cells were distributed exponentially in
radius, with the innermost cell located inside the BH horizon at
$r=1.85\rg$. The polar cells were uniformly distributed.

All the simulations except the high accretion rate $\ttt{Rad4}$
were initialized with the same torus in hydrostatic equilibrium. The
parameters of the torus were set following \cite{narayan+adaf}. In
particular, the inner edge of the torus was located at $10\rg$, which, for the given
angular momentum profile, resulted in a hot and geometrically thick
structure.\footnote{We note that it is essential to start simulations
  of optically thin flows with a hot and thick torus. Otherwise,
  the gas may not have enough time (which is limited by the viscous
  timescale) to reach the correct local equilibrium temperature (equal roughly to the virial
  temperature) before falling into the BH. This would not be an issue
  if we could start the gas far from the BH, as in real systems in
  nature. However, because of computational limitations, the accreting
  gas in simulations has to be initialized relatively close to the BH.} The torus was threaded with multiple loops of a weak
(contributing at most to 3\% of the gas pressure)
magnetic field with loop properties identical to the ones adopted in
\cite{narayan+adaf}. The density of the gas in the torus was chosen
by adjusting the torus entropy parameter \citep{penna+limotorus} 
to provide the required accretion rate. The torus was initially surrounded by a
a non-rotating atmosphere of negligible mass.

At the onset of the simulations we set the electron and ion temperatures
equal to the gas temperature (which may not be a good assumption if
Coulomb coupling is ineffective and the particles are not evolved long
enough to forget the initial state). The radiation field both inside and
outside the torus was initiated with a~negligible energy density and
radiation temperature of $10^5 \rm\, K$.

To start the higher accretion rate simulation \ttt{Rad4} we chose a~different approach and took a very early stage of simulation
\ttt{Rad8}, rescaled the density and magnetic field pressure up by
four orders of magnitude (keeping all the temperatures
constant), and evolved it from there. Such an approach is advantageous for higher accretion
rates, where the large optical depth requires non-negligible radiation
pressure inside the torus, and starting with a virtually empty
radiation field would not be a good approximation.

\begin{table*}
\begin{center}
\caption{Model parameters}
\label{t.models}
\begin{tabular}{llccccc}
\hline
\hline
Name &\hspace{.45cm}  &
$M/M_{\odot}$ & $\dot M_{\rm BH}/\dot M_{\rm Edd}$ &
$L/L_{\rm Edd}$ & adiabatic index \\
\hline
\hline
\texttt{Rad8} && 10 & $4.4\times 10^{-8}$ & $2.1\times 10^{-9}$ & consistent \\
\texttt{Rad8SMBH} && $4\times10^6$ & $1.9\times 10^{-8}$ & $4.6\times 10^{-10}$ & consistent \\
\texttt{Rad4} && 10 & $4.5\times 10^{-4}$& $3.8\times 10^{-4}$  & consistent \\
\texttt{Hd8} && 10 & $4.4\times 10^{-8}$ & N/A & consistent \\
\texttt{Hd8fg} && 10 & $3.2\times 10^{-8}$ & N/A & $\gamma_{\rm gas}=5/3$,
$\gamma_{\rm int\, e}=4/3$ \\
\hline
\hline
\multicolumn{6}{l}{Other parameters: 
  $a_*=0$, $r_{\rm in}=1.85\rg$, $r_{\rm out}=1000\rg$. Resolution:384x384x1}\\
\multicolumn{6}{l}{Accretion rate and luminosity measured at $r=20\rg$.}\\
\end{tabular}
\end{center}
\end{table*}

\subsection{Properties of the models}
\label{s.properties}

\subsubsection{Fiducial model}

In Fig.~\ref{f.d250} we show properties of the fiducial model
\texttt{Rad8} corresponding to a $10\msun$ BH accreting at a very low
accretion rate of $\sim 4\times 10^{-8}\Medd$, and emitting $\sim
2\times 10^{-9}L_{\rm Edd}$\footnote{This is manifestly a radiative
  inefficient accretion flow (or ADAF), since a radiatively efficient
  flow at this accretion rate would have a luminosity of $\sim 4\times
  10^{-8}L_{\rm Edd}$, i.e. 20 times larger.}. Such a system can
represent a transient BH X-ray
binary in deep quiescence, in between the outbursts.
We chose such a low accretion rate for the fiducial model
to verify whether radiative cooling is important for accretion rates
comparable with those characterizing the accretion on Sgr A*. 
The impact of the accretion rate, BH mass, and
radiative feedback is studied via the remaining simulations.

The left and right halves of Fig.~\ref{f.d250} show for the fiducial model the distributions
in the poloidal plane obtained from a snapshot (corresponding to
$t=15000\tg$) and time-averaged
over $t=10000\tg$ -- $20000\tg$, respectively. The top-most panel shows the distribution of gas
density (colors) and the velocity field (streamlines). The disk is evidently thick and turbulent.
On average, the gas flows towards the BH in
the equatorial region. Outflows emerge from the
surface layers outside $r\approx 20\rg$.  

The second panel shows the magnetization parameter $\beta=p_{\rm
  tot}/p_{\rm mag}$ with magnetic field lines overplotted (blue
and red contours correspond to clockwise- and counter-clockwise
poloidal magnetic field). Note that the instantaneous magnetic field
is turbulent. 
The amount of the magnetic flux that has accumulated at
the horizon is far below the magnetically arrested limit
\citep{narayan+mad,sasha+madjets} 
because of the initial magnetic
field setup of alternating polarity which causes the loops
subsequently accreted onto the BH to reconnect with the magnetic flux
already present in the inner region. 
 The polar region, where the gas
density is the lowest, is highly magnetized ($\beta<1$). The bulk of
the disk, on the other hand, is dominated by gas pressure
($\beta>1$). The magnetization level at the equatorial plane increases
towards the BH from $\beta\approx 10$ at $30\rg$ to $\beta\approx 5$
at $15 \rg$.  The global picture of the gas dynamics resembles well
the optically thin accretion flows obtained numerically in past
simulations \citep[e.g.,][]{narayan+adaf}.

The third panel in Fig.~\ref{f.d250} shows the electron temperature
obtained by consistently evolving electrons during the
simulation. In the fiducial model the electrons were dragged with the gas,
underwent adiabatic evolution, were affected by a fraction of the viscous
heating (Section~\ref{s.viscous}), exchanged energy with the ions through 
Coulomb interactions (Section~\ref{s.coulomb}), and cooled by
emitting radiation (Section~\ref{s.Gi}). As will turn out later, for
such a low accretion rate and gas density, radiative feedback and
Coulomb coupling are negligible. The left half of the third panel
shows the instantaneous distribution of electron temperature. Electrons
are coldest far from the BH and in the bulk of the disk -- at
$r\approx 50$ their temperature does not exceed $10^{10}\rm\, K$. They heat
up on the way towards the BH, mainly because of adiabatic
compression. The viscous heating is weak near the equatorial plane
where the magnetic pressure is sub-dominant and most of the heating
goes into the ions. In the polar region, on the contrary, the magnetic field
is strong and as a result most of the viscous heating goes into the
electrons (as per our heating subgrid model, Eqs.~\ref{e.deltae}\,\&\,\ref{e.deltae2}). In the instantaneous temperature profile, electron filaments are evident 
at $\sim 10^{11.5}\rm\, K$ overlapping with the regions
of 
strongest magnetization. Such hot electrons enter the polar region and strongly
contribute to the time-averaged properties (shown in the right
half of the panel), covering the entire polar region with hot
electrons with an average temperature
of $10^{11}\rm\, K$. 

The fourth panel in Fig.~\ref{f.d250} shows the electron to ion
temperature ratio. The two species compete in how much dissipative
heating goes into each. In regions where the magnetic pressure
dominates over the gas pressure (in our case this corresponds to the
polar region), most of the heating goes into the electrons
(Eqs.~\ref{e.deltae}\,\&\,\ref{e.deltae2}). In the opposite case,
where the magnetic pressure is sub-dominant (in the bulk of the disk),
the ions are heated more efficiently. In addition to viscous heating,
electrons and ions exchange energy in Coulomb collisions. We also
account for radiative cooling which affects only electrons. In the
case of the fiducial model, the electron/ion temperature balance is
determined mostly by the sub-grid viscous heating prescription
(Eq.~\ref{e.deltae}, Section~\ref{s.Tediscussion}) which determines
the fraction of heating going into each of the species as a function
of their temperatures and local magnetization. In regions dominated by
the magnetic pressure, electrons are heated more efficiently than
ions. If gas pressure dominates (e.g., near the equatorial plane),
ions receive most of the dissipated heat. As a result, the two regions
can be distinguished. In the bulk of the disk, where magnetic pressure
is sub-dominant and reflects the saturated state of MRI, ions are
hotter than electrons ($T_{\rm e}/T_{\rm i}\approx
0.2$ near $20\rg$). In the polar region, on the contrary, electrons carry more
energy per particle and $T_{\rm e}/T_{\rm i}\approx 2$. The boundary
between the two regions agrees well with the surface where the
magnetic pressure equals the gas pressure, $p_{\rm mag}=p_{\rm gas}$
(compare the contour lines in the right half of the fourth panel in
Fig.~\ref{f.d250}).

The electron temperature exceeds $10^{9.5}\rm K$ virtually everywhere
inside $r=50\rg$. Therefore, electrons are relativistic and their
equation of state can be characterized by the adiabatic index,
$\gamma_{\rm int\,e}=4/3$. Ions, on the other hand, never exceed
$10^{11.5}\rm K$, which, keeping in mind their much larger mass, makes
them non-relativistic ($\gamma_{\rm int\, i}\approx 5/3$)
everywhere. The effective adiabatic index of the electron and ion
mixture depends on both temperatures (Eq.~\ref{e.gammagas}) and is not
dominated by the more massive ions (ions and electrons at the same
temperature carry the same energy per particle).  The fifth panel in
Fig.~\ref{f.d250} shows the effective adiabatic index of the total
gas, $\gamma_{\rm gas}$. In the polar region, where electrons are the
hottest and most relativistic, the electron/ion mixture effectively
follows $\gamma_{\rm gas}\approx4/3$. In the bulk of the disk, on the
other hand, where the electrons are only moderately relativistic, the
adiabatic index of the gas is significantly higher, $\gamma_{\rm
  gas}\approx 1.55$. Therefore, assuming a single fixed value for the
gas adiabatic index for the whole domain is probably not a good
approximation.

The methods described in this paper allow for simulating the parallel
evolution of the gas and the radiation field. The latter is generated
self-consistently by the gas which cools by emitting bremsstrahlung
and synchrotron radiation. Subsequently, the emitted photons only
interact with the gas they are penetrating if the gas's optical depth
is large enough, which can lead to the exchange of energy and
momentum.  In the two bottom-most panels of Fig.~\ref{f.d250} we show
the properties of the radiation field generated by gas in the fiducial
model \texttt{Rad8} ($10\msun$, non-rotating BH, $\sim 3\times
10^{-8}\Medd$). The second panel from the bottom shows the magnitude
of the radiative flux (colors) and its direction in the poloidal plane
(streamlines). Radiation propagates outward almost isotropically. This
reflects the fact that the gas is optically thin and its distribution
does not affect the propagation of photons. Most of the flux comes
from the innermost, hottest, and densest region, where synchrotron
emission is most efficient. The integrated radiation flux gives a
luminosity on the order of $\sim 10^{-9}L_{\rm Edd}$.

The last panel in Fig.~\ref{f.d250} shows the characteristic
temperature of the emitted radiation. At such low accretion rates,
virtually all of the photons are synchrotron photons and carry energies
corresponding to frequencies $10^{13}\text{ Hz}$ -- $10^{14}\text{ Hz}$ (for a
$10\msun$ BH). The angular
distribution of photon energies is not isotropic, in contrast to the
radiative flux distribution. The most energetic photons are generated in
the polar region (because of highest electron temperatures there) and
tend to propagate upward. Colder electrons, closer to the equatorial
plane, generate less energetic photons. Radiative
postprocessing is required to obtain more detailed information on the emerging
electromagnetic spectrum.

\begin{figure*}
\begin{flushleft}
 \hspace{1.8cm}\texttt{Hd8} \hspace{2.2cm}\texttt{Hd8fg} \hspace{2.6cm}\texttt{\textbf{Rad8}}
\hspace{2.2cm}\texttt{Rad8SMBH} \hspace{2.2cm}\texttt{Rad4} \\
 \hspace{1.4cm}$4.4\times 10^{-8}\dot M_{\rm Edd}$  \hspace{1.05cm}$3.2\times 10^{-8}\dot
M_{\rm Edd}$  \hspace{1.1cm}$4.4\times 10^{-8}\dot M_{\rm Edd}$
\hspace{1.45cm}$1.9\times 10^{-8}\dot M_{\rm Edd}$  \hspace{1.35cm}$\sim
3.5\times 10^{-4}\dot M_{\rm Edd}$  \hspace{1.45cm}\\
 \hspace{1.8cm}$10M_{\odot}$  \hspace{1.3cm}$10M_{\odot}$,
$\gamma_{\rm gas}=\frac53,\,\gamma_{\rm int\,e}=\frac43$
\hspace{1.0cm}$10M_{\odot}$  \hspace{2.15cm}$4\times10^{6}M_{\odot}$
\hspace{2.05cm}$10M_{\odot}$  \hspace{2.25cm}\\
\end{flushleft}
 \includegraphics[width=1.\textwidth]{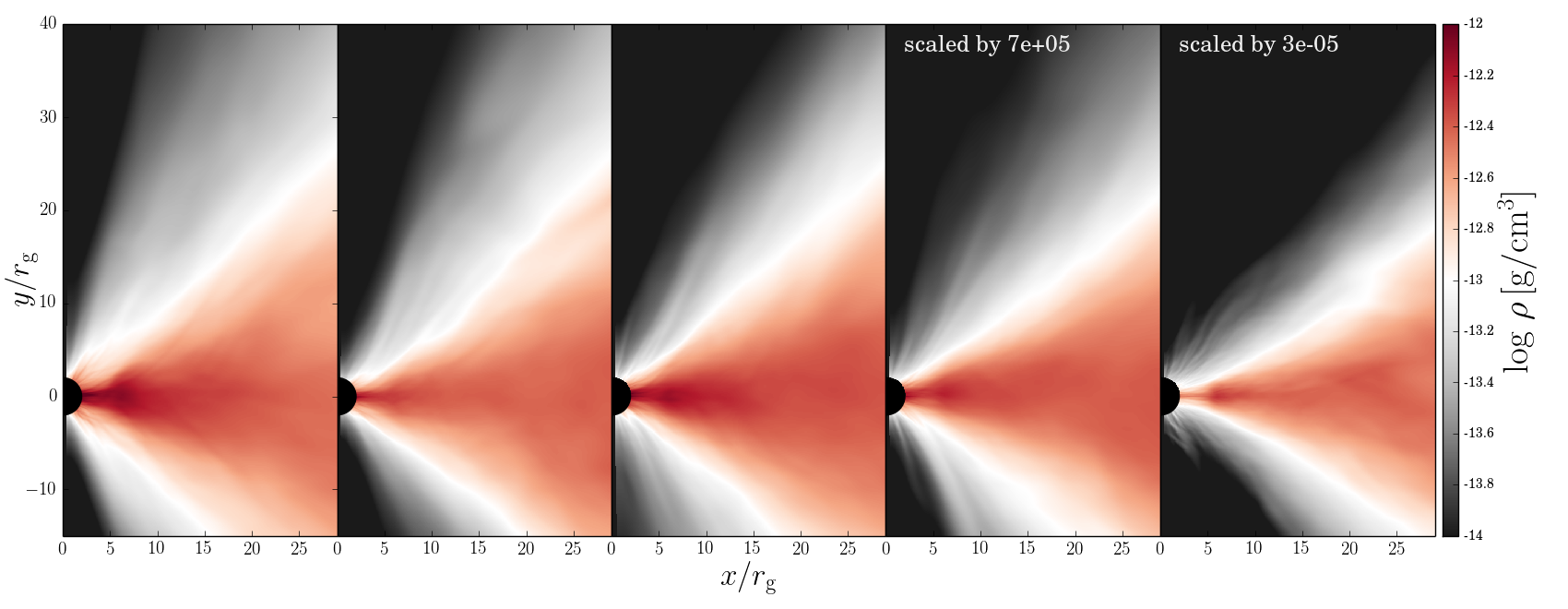}\\\vspace{-.55cm}
 \includegraphics[width=1.\textwidth]{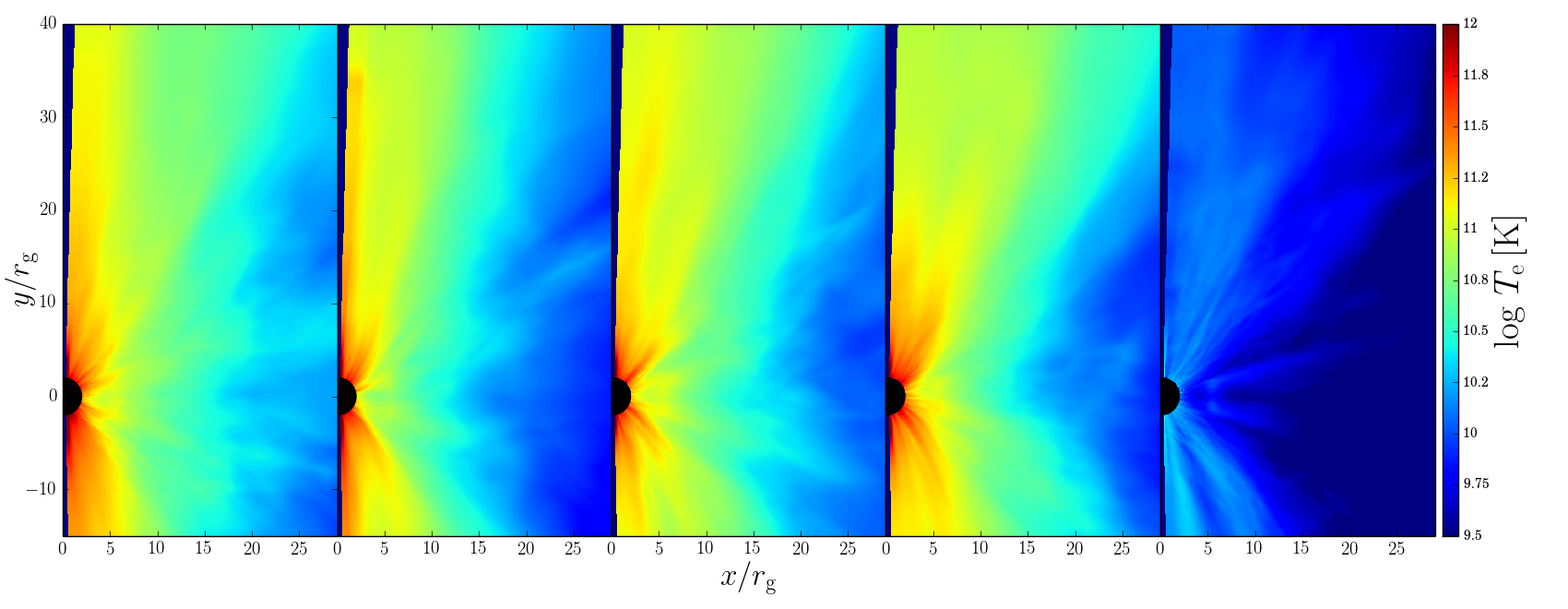}\\\vspace{-.55cm}
 \includegraphics[width=1.\textwidth]{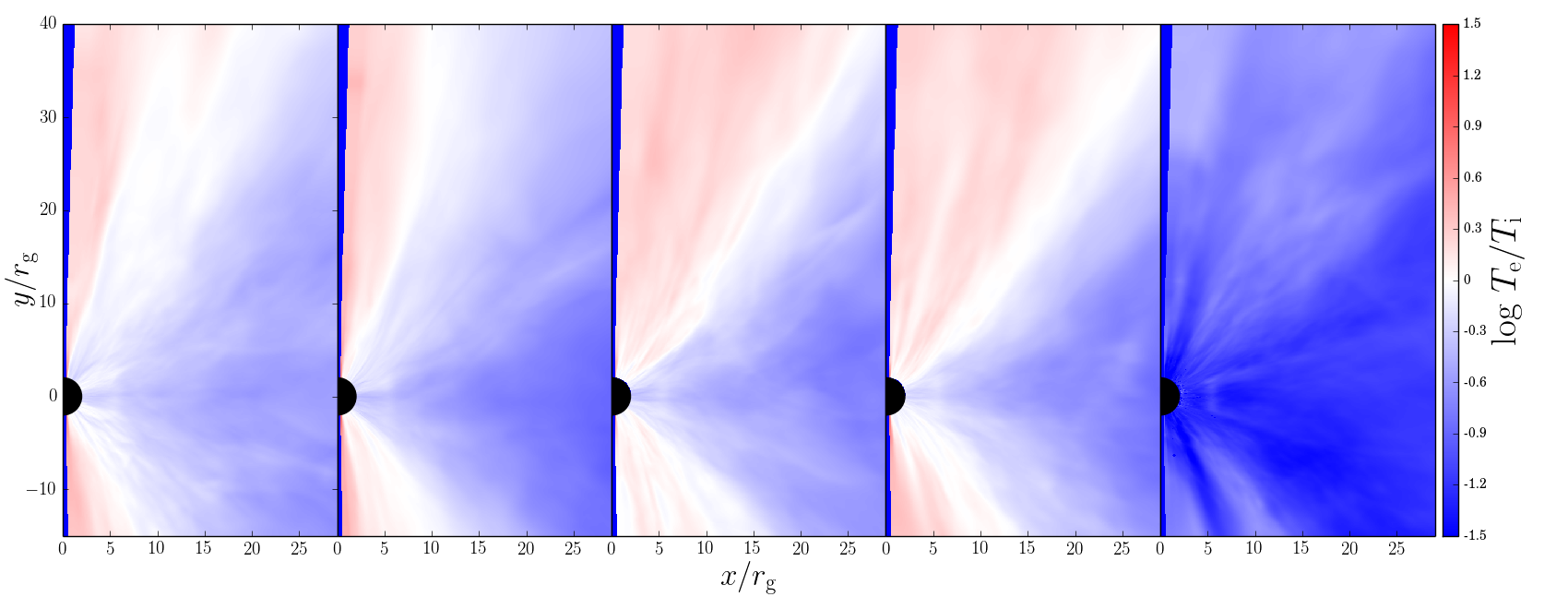}
 \caption{Time-averaged distribution of logarithms of density (top), electron temperature (middle)
   and electron to ion temperature ratio (bottom) for 
   (left to right) models \texttt{Hd8}, \texttt{Hd8fg}, \texttt{Rad8},
   \texttt{Rad8SMBH}, and \texttt{Rad4}.  }
 \label{f.multi1}
\end{figure*}

\begin{figure*}
\texttt{Rad8}
\hspace{2.4cm}\texttt{Rad8SMBH} \hspace{3.cm}\texttt{Rad4}\\
$4.4\times 10^{-8}\dot M_{\rm Edd}$  \hspace{1.45cm}$1.9\times 10^{-8}\dot M_{\rm Edd}$  \hspace{1.845cm}$4.5\times 10^{-4}\dot M_{\rm Edd}$ \\
$10M_{\odot}$  \hspace{2.15cm}$4\times10^{6}M_{\odot}$  \hspace{2.85cm}$10M_{\odot}$ \\
 \includegraphics[width=.7\textwidth]{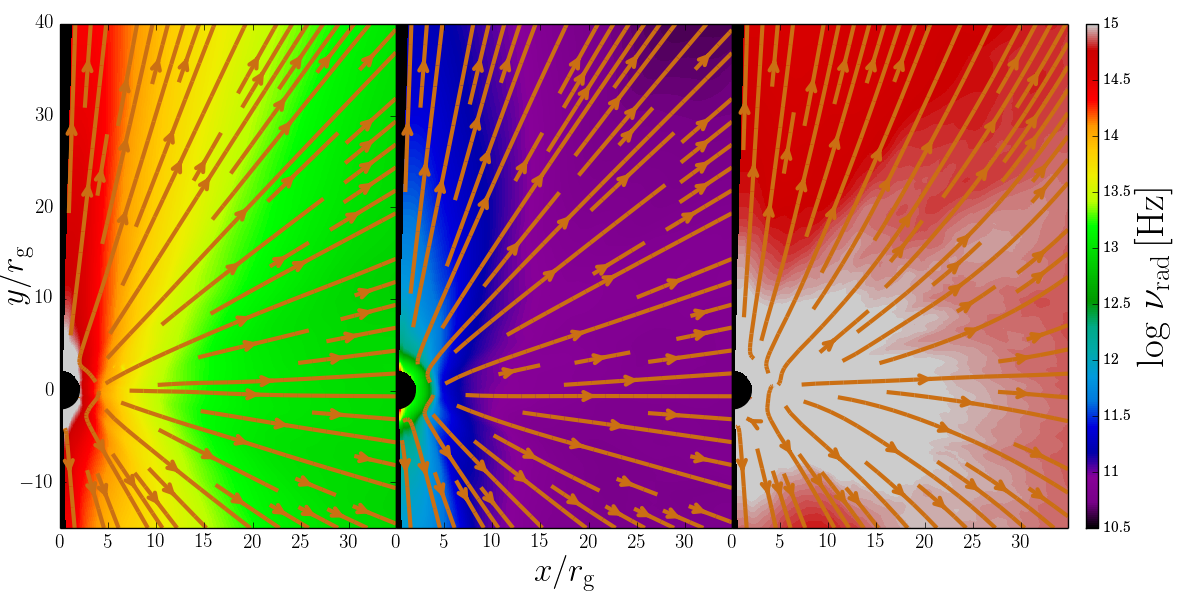}
 \caption{Characteristic frequency of emitted radiation for models
   (left to right) \texttt{Rad8}, \texttt{Rad8SMBH}, and \texttt{Rad4}.
  }
 \label{f.multi3}
\end{figure*}

\subsubsection{Parameter study}
\label{s.parameter}

In this Section we study how the choice of the adiabatic index,
radiative feedback, BH mass, and accretion rate affect the properties
of low-luminosity accretion flows.

Figure~\ref{f.multi1} compares the gas density (top), electron
temperature (middle), and electron to ion temperature ratio (bottom
panels) distributions for the five models we have simulated. 

Let us first compare model \ttt{Hd8} (first column) with the fiducial
simulation \ttt{Rad8} (third column). The former was initiated with
exactly the same equilibrium torus as the latter, and adopted the same
generalized entropy prescription, which allows for a consistent
adjustment of the adiabatic index in the course of the simulation. In
contrast to the fiducial model, however, it did not evolve the
radiation field, so that radiative cooling by electrons was
neglected. As a comparison of the panels of density, electron temperature, and 
temperature ratio in  Fig.~\ref{f.multi1}
shows, the properties of the two models are virtually
indistinguishable. This shows that, at such a low accretion rate
($\sim 4\times 10^{-8}\Medd$), the impact of radiation on the accretion flow
properties is negligible. 

Model \texttt{Hd8fg} (second column in Fig.~\ref{f.multi1}) was
initiated from the same initial state, but the adiabatic indices of
the total gas and the electrons were not adjusted according to the
local ion and electron temperatures. Instead, the respective adiabatic
indices were fixed at $\gamma_{\textrm{gas}}=5/3$ and $\gamma_{\rm
  int\, e}=4/3$.  In addition, the dissipation was identified by
comparing adiabatic and non-adiabatic evolution of the total gas, not
the particular species.  This approach turns out to give qualitatively
similar results for the properties of the accretion flow. In
particular, the electron temperatures are quite close. Only the
density distributions show slight differences, as may be expected for
gas evolving under different equations of state.

The fourth column in Fig.~\ref{f.multi1} corresponds  to model
\ttt{Rad8SMBH} which is similar to the fiducial model \ttt{Rad8}, but the BH mass is different
and roughly matches that of the supermassive BH in the center of our
galaxy. The initial torus of gas in \ttt{Rad8SMBH}  had the same
dynamical properties as in \ttt{Rad8},
but the density of the gas was scaled
to provide a comparable accretion rate in Eddington units. The top panels
in columns 3 and 4 compare the gas density distributions. In the case of model
\ttt{Rad8SMBH}, the real densities were rescaled by
the ratio of the BH masses and the accretion rates, so that the color
scale can be matched between the panels. Both the density
distribution and the electron/ion temperatures (shown in the second
and third rows) almost exactly correspond to each other. This, once
again, proves that radiative feedback does not affect accretion
flows at such low accretion rates, independent of the BH mass.

The properties of the radiation field, however, are different in
models \ttt{Rad8} and \ttt{Rad8SMBH}. The more massive BH in \ttt{Rad8SMBH} results in 
a~larger radiative luminosity (which still corresponds to roughly the
same Eddington fraction), and a different electromagnetic spectrum. For
low-luminosity accretion flows, as discussed so far, radiation is
dominated  by synchrotron emission. The characteristic frequency of
the emitted radiation (Eq.~\ref{e.nusyn}) depends on the strength of
the magnetic field and the square of the electron temperature. While the
latter is independent of BH mass (as long as radiative cooling is
weak), the strength of the magnetic field scales
as $1/\sqrt{M_{\rm BH}}$. Thus, one expects the peak frequency in the electromagnetic spectrum for
  model \ttt{Rad8SMBH} to be roughly two orders of
  magnitude lower. As Fig.~\ref{f.multi3} shows, this is, indeed,
  the case. The fiducial model ($M_{\rm BH}=10\msun$, left panel) emits
  radiation with characteristic frequencies around $10^{14}\,\rm Hz$ and
  $10^{13}\,\rm Hz$ near the axis and in the equatorial plane,
  respectively. Model \ttt{Rad8SMBH} ($M_{\rm BH}=4\times 10^6
  \msun$, middle panel), on the other hand, produces less energetic photons and the
  spectrum is expected to peak near $10^{11.5}$ and $10^{10.5}\,\rm
  Hz$, respectively. It is remarkable that these are roughly the frequencies
  where the synchrotron emission from Sgr A* peaks
  \citep{yuan+03}, which proves that our algorithm, despite
  using a grey approximation, is able to track the temperature of
  radiation reasonably well.

Finally, we compare the fiducial model \ttt{Rad8} with model
\ttt{Rad4}, which was designed to correspond to an accretion rate
roughly 4 orders of magnitude larger, i.e., $\sim 10^{-4}\Medd$. This
model was initialized from an early stage of the fiducial simulation,
after rescaling the density up by four orders of magnitude (keeping
magnetic to gas pressure ratio and all the temperatures fixed). Once
the density was rescaled, the radiative cooling immediately started to
affect the electrons. This effect is clearly seen in the rightmost
column in Fig.~\ref{f.multi1}. There are noticeable differences
between model \ttt{Rad4} and all the other simulations discussed so
far -- the disk is confined to a thinner region near the equatorial
plane, and the gas is much colder than in the previous, low accretion
rate models. The temperature at the equatorial plane is as low as
$10^9\rm K$ at $15\rg$.  These distinct properties result from thermal
cooling of an initially thick and hot accretion disk.

\section{Discussion}
\label{s.discussion}

\subsection{Electron temperatures}
\label{s.Tediscussion}

\begin{figure}
 \includegraphics[width=1.\columnwidth]{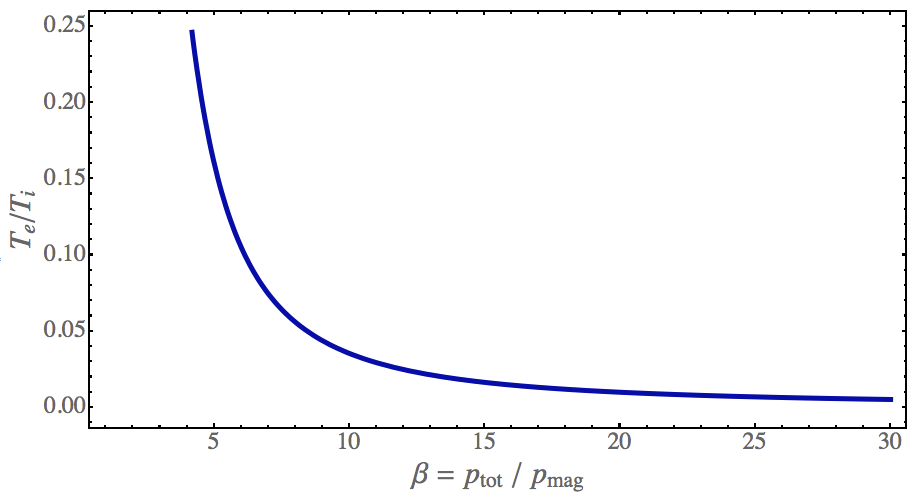}
  \caption{Electron to ion temperature ratio corresponding to the
    equilibrium state (eq.~\ref{eq:teti}) for given magnetization
    parameter $\beta$ as implied by the Howes (2010) heating
    prescription. See Section~\ref{s.Tediscussion} for details.}
 \label{fig:teti_saturated}
\end{figure}

The final electron temperature depends on the initial temperature
(both in the temporal and spatial sense), and the amount of
dissipation that the electrons undergo. If the amount of energy added
by dissipation is much smaller than the initial internal energy, then
the final temperature of the electrons will differ from the original
only by the adiabatic compression term. If, on the contrary, the
amount of the dissipation is much larger than the original energy
content of electrons then the initial state is quickly forgotten.

Our simulations of low-luminosity accretion flows show that the latter
is the case. In particular, the gas density in the inner region is
very close to constant (Fig.~\ref{f.multi1}), and therefore adiabatic
compression does not modify the thermal energy as the gas moves
towards the BH. However, the gas temperature shows a steep $\sim1/R$
variation with radius, which implies that the thermal energy must
increase ten-fold when crossing a decade in radius. Thus, the initial
energy content is quickly forgotten and it is the dissipated energy
(magnetic or kinetic) that heats up the gas as it flows in.  This is
in contrast to the properties of self-similar, $\gamma=5/3$, models of
ADAFs \citep{narayanyi-95} which predict no dissipation at all.

This fact has significant consequences. In particular, assuming
$\gamma_{\rm e}=4/3$ and $\gamma_{\rm i}=5/3$, in the limit of
dissipation dominating, one finds that the final electron to ion
temperature ratio is given by, \be \frac{T_{\rm e}}{T_{\rm
    i}}=\frac12 \frac{\delta_{\rm e}}{1-\delta_{\rm e}}, \ee where
$\delta_{\rm e}$ is the fraction of the dissipated energy going into
the electrons.

The heating prescription adopted in this work
(Section~\ref{s.viscous}) gives $\delta_e$ as a function of the
magnetization parameter, $\beta=p_{\rm tot}/p_{\rm mag}$, and the
electron to ion temperature ratio, $T_{\rm e}/T_{\rm i}$. There is an
equilibrium value of this ratio for which applying viscous heating
does not modify the temperature distribution. This equilibrium ratio
is given through,
\be \left(\frac{T_{\rm e}}{T_{\rm i}}\right)_{\rm eq}=\frac12
\frac{\delta_{\rm e}\left(\left(T_{\rm e}/T_{\rm i}\right)_{\rm
    eq},\beta\right)}{1-\delta_{\rm e}\left(\left(T_{\rm e}/T_{\rm
    i}\right)_{\rm eq},\beta\right)}. \label{eq:teti} \ee
Figure~\ref{fig:teti_saturated} shows the equilibrium temperature
ratio as a function of the magnetization parameter. It is evident that
this ratio depends strongly on the gas magnetization parameter,
e.g., it equals $0.066$ and $0.243$ for $\beta=10$ and $5$,
respectively. The magnetization of our fiducial model near $r=20\rg$
is close to $\beta=5$, and the value of $\left(T_{\rm e}/T_{\rm
  i}\right)_{\rm eq}$ predicted for this magnetization is indeed close
to the one obtained in the simulation ($0.2$, see
Fig.~\ref{f.multi1}).

Shearing box simulation with no vertical magnetic field have shown
that the magnetic field saturates near $\beta=10$. However, inclusion
of even a small amount of net vertical flux can drive this parameter
down substantially \citep{xuening+13}. Such a behavior has been shown
to affect global simulations as well \citep[e.g.,][]{narayan+mad,
  sadowski-thin}.

One should keep in mind the implicit relation betwen magnetization and
electron temperature (eq.~\ref{eq:teti}) when interpreting and
comparing results of simulations initiated with different
configurations of the magnetic field. Likewise, this strong dependence
could be used in the future to infer the properties of the magnetic
field in accretion disks from their observed radiative properties.

We should also mention that, ultimately, three-dimensional simulations
will be needed to verify that the axisymmetric, dynamo-based models
presented here capture the heating rates accurately.

\subsection{Implemented and not implemented physics}

Our code KORAL evolves the accretion flow under the magnetohydrodynamical 
(MHD) approximation, i.e., we treat electrons and ions as fluids
characterized by their own, isotropic, pressures. In reality,
electrons and ions are particles moving in self-generated magnetic and
electromagnetic fields, and are expected to undergo plasma
instabilities. Under the adopted MHD
approximation we effectively assume that all these instabilities
saturate such that the total gas and its individual species (electrons
and ions) can be reasonably well described by an isotropic pressure
and adiabatic indices, with appropriate distinctions between
$\gamma_{\rm CV}$ and $\gamma_{\rm int}$ (see Appendix~\ref{ap.entropy}).
This,
however, does not have to be the case \citep[see][and voluminous
literature on collisionless plasmas]{sharma+07}. Whether or not
this fluid approximation works has to be verified in the future by
using a microscale particle approach in the global context of an
accretion disk.

In grid-based codes like KORAL, kinetic and magnetic energy is
dissipated into thermal energy on the grid scale through
numerical dissipation. As long as energy is conserved, such an approach gives
the proper dissipative (viscous) heating of the gas. However, if one
wants to distinguish electrons from ions, it becomes crucial to know
how much of the dissipated energy goes into each species, which is
determined on microscales, far below the resolution. In
this work, following \cite{ressler+15}, we adopted the sub-grid prescription of \cite{howes+10}
which is based on theoretical models of the dissipation of MHD
turbulence in almost collisionless 
plasmas and to some extent
validated by numerical experiments. It is still,
however, only an approximation and could be avoided once hybrid
algorithms, combining the particle and global frameworks, become
feasible. What is more, one should in principle allow also for
non-thermal, relativistic electron population, which may not change
the overall dynamics
 of the two-phase accretion flow, but may be relevant to the generated
radiation.

Energy can be transported in many ways. We account for a number of them, including,
in particular, radiative transport and the exchange of energy between gas
and the radiation field. We ignore, however, anisotropic heat
conduction along magnetic field lines. The effect of this mechanism on
optically thin accretion flows has recently been studied by \cite{ressler+15},
\cite{chandra+15}, and \cite{foucart+16}. \cite{ressler+15}, in particular,
have shown that 
the anisotropic heat conduction contributes to at most $20\%$
of the isotropic heat flux. The low efficiency of
the anisotropic conduction results from the fact that the magnetic field
in accretion disks 
is predominantly toroidal and there are no significant temperature
gradients in that direction. The same authors, however, claim that
it might be better to set the conduction parameter to $\alpha_{\rm e}\approx 10$ as the resulting heat flux corresponds better to that expected
in collisionless plasmas.

Finally, we mention a set of caveats not specific to the algorithm
presented in this work, but applicable also to standard, single-fluid,
radiative simulations. We adopt the ideal-MHD approximation, i.e.,
we implicitly assume that the resistivity is so low that there is no
electric field in the comoving frame of the gas. Such an assumption is
reasonable as long as we deal with a fully ionized, hot gas, which is the
case for astrophysical BH accretion disks. We do not include
explicitly the resistive term. However, magnetic reconnection
still takes place due to numerical, grid-scale resistivity. One may
wonder if numerical resistivity impacts the rate at which the magnetic
field is advected onto the BH which depends on the effective Prandl
number. In most MHD accretion flows
 the Prandtl number is dominated by the turbulent component and
 therefore not sensitive to microscopic (neglected by us) resistivity
 \citep{guangammie-09}. 

We treat the radiation transport in an approximate way. First, we
solve it in the grey approximation, imposing a diluted black body
shape for the spectrum, though we do use consistent grey opacities for bremsstrahlung and synchrotron
processes. As long as the spectrum is dominated by a single component
(either synchrotron, Compton or free-free), our approximation is
reasonable. However, one should be careful when two of the emission
mechanisms contribute to the spectrum at a similar level. Our
assumption of a diluted black body shape would mix the two components
into one and the resulting opacities would not be
correct. Fortunately, in low-luminosity accretion flows synchrotron
emission always dominates, and one may expect that the transition to
bremsstrahlung and Comptonization with increasing accretion rate would be rapid enough not
to cause any trouble. Finally, we adopt the M1 closure scheme
\citep{levermore84,sadowski+koral2} which describes the shape of the
local radiation field with just four numbers: its energy density and
three spatial components of momentum. Obviously, it has its
limitations, especially when handling multiple sources of
radiation. This is not, however, a serious concern in optically thin,
low-luminosity accretion flows, where most of the emission
comes from the inner region and propagates isotropically outward (see
the last-but-one panel in Fig.~\ref{f.d250}) which the
M1 closure can handle  reasonably well. 

\subsection{Uncertainities and limits}

In this work we have presented five simulations designed to help
assess the importance of particular factors on accretion flow
properties. We studied the impact of radiative cooling at two, very
different, accretion rates, the impact of fixing adiabatic indices to
constant values, and finally, the importance of the BH mass. This set,
however, does not exhaust the parameter space. Most importantly, the
properties of an accretion flow, especially in the polar region, are
known to depend on the large scale topology of the magnetic field. In
particular, if the magnetic field exhibits uniform polarity on long
timescales, then the magnetic field deposited on the BH and
accumulated on the axis may dynamically affect the flow dynamics
forming a so-called magnetically arrested disk
\citep[MAD,][]{narayan+mad,sasha+madjets}. For rotating BHs this can
result in efficient
jet production. In a similar way, the magnetization of gas,
and in turn, the heating of electrons and their radiative properties,
would be affected. All the simulations discussed in this work were
initialized with the same magnetic field setup. We plan to study other 
initial field configurations in the future.

The most exciting application of the numerical methods developed here is
to study in detail the transition from totally optically thin
accretion flows towards geometrically thin but optically thick
disks. One has to be aware, however, of the  difficulties related to
simulating the latter. The standard thin disk model \cite{ss73} predicts that at
accretion rates on the order of $10^{-2}\Medd$, the expected disk
thickness is $H/R\sim 0.01$. To resolve the fastest growing MRI mode,
this disk thickness has to be resolved by at least a few dozens of
cells. This introduces a very demanding criterion for vertical
resolution and time step. In addition, geometrically thin disks
evolve on very long
viscous time scales, $t_{\rm vis}\approx (R/H)^2 /\alpha \Omega_{\rm 
  K}$. For obvious reasons, the duration of simulations is limited,
and therefore one cannot hope to obtain an equilibrium solution for a~thin 
disk covering a large range of radii. Last but not least, there is the issue of
thermal stability which kicks in when portions of a~thin disk become
radiation pressure dominated ($\sim 0.1\Medd$), which seems to require
a~strong magnetic field for stabilization \citep{sadowski-thin}. All
these factors will require special attention when studying the
transition in detail.

\subsection{Comparison with previous studies}

In this Section we discuss the relation of our work to some of the 
related works published in recent years.

\subsubsection{Ressler et al. (2015)}

In \cite{ressler+15} the authors introduced a method of evolving
electrons in GR magnetohydrodynamical (MHD) simulations. We follow
their approach and adopt the same sub-grid prescription
\citep{howes+10} for the fraction of dissipation going into
electrons. The authors account for anisotropic heat conduction along
magnetic field lines which we ignore, as they found it is always
sub-dominant. Their approach, however, adopts a number of
simplifications which we have avoided. Firstly, their electrons are
decoupled from the evolution of the gas, i.e., they evolve electrons
only during the postprocessing stage and the electrons do not affect
the dynamics of the gas. Secondly, they assumed fixed adiabatic
indices of gas and electrons throughout the simulations. This last
assumption is not generally accurate -- the effective adiabatic index
of the gas cannot be approximated by $\gamma_{\rm gas}=5/3$, instead,
a consistent value, dependent on the instantaneous and local electron
and ion temperatures, should be used. However, in practice, the
difference is not large, and simulations run with fixed $\gamma_{\rm
  gas}$ do produce qualitatively similar results as simulations that
use a more accurate variable $\gamma$, at least in the case of hot
accretion flows with low $\dot{M}$. \cite{ressler+15} also neglect
radiative cooling and feedback, as well as Coulomb coupling. This is a
valid approximation for accretion rates much less than $10^{-8}\Medd$,
but it cannot be used for rates much larger than this.

\subsubsection{Dibi et al. (2012)}

The critical accretion rate above which radiation cooling becomes
important and affects properties of the accretion flow was the topic of
another study -- \cite{dibi+12}. The authors implemented
simplified radiative transfer in GR MHD simulations \citep[following][]{esin+96}, and, in this way,
were able for the first time to include radiative cooling. They found
that, as long as the accretion rate is below $6\times 10^{-8}\Medd$
(in our units, note the different definition of the Eddington
accretion rate adopted in their work), radiative effects are
negligible. Our fiducial model (\ttt{Rad8}), accreting at a level of
$3\times 10^{-8}\Medd$, is unaffected by radiative cooling, in
contrast to the high accretion rate model (\ttt{Rad4}), showing
significant radiative feedback. Both conclusions are consistent with the
findings of \cite{dibi+12}. We will determine the exact value of the critical
accretion rate in a future work.

Let us, however, point out differences between the two works. In terms of
the electron evolution and radiative transfer, our work supersedes
\cite{dibi+12} in all aspects. Instead of adopting, as they did, a local cooling
based on local gas properties and an estimate of the temperature
scale height, we evolve the radiation field simultaneously and in the
whole domain (although using an approximated closure scheme). Instead
of fixing the electron to ion temperature ratio throughout the
computational box,
we calculate the two temperatures consistently. We also do not fix the
adiabatic index of the gas. We will have to perform a more extensive
parameter study to assess how well the
critical accretion rate obtained with our code agrees with the
predictions of \cite{dibi+12}.

\subsubsection{Wu et al. (2016)}

Numerical methods for approximate treatment of radiation similar
  to \cite{dibi+12} have
been adopted recently in \cite{wu+16}, although the authors used a
Newtonian MHD code. Similarly to \cite{dibi+12}, the electron
temperature was arbitrarily prescribed. The main point of interest of
\cite{wu+16} was the state transition taking place at accretion rates
of the order $\sim 10^{-2}\Medd$. The authors have identified the 
formation of cold and dense clumpy/filamentary structures embedded
within the hot gas, which, when the accretion rate becomes
sufficiently high, gradually merge
and settle down onto the mid-plane. Their findings are in good
agreement with what we found for our high-accretion rate model
\texttt{Rad4}. We point out, however, that our methods supersede the
approach adopted both in \cite{dibi+12} and \cite{wu+16}, allowing for
simulating the transitional regime more reliably. We
plan to verify findings of these works in future studies.

\subsubsection{Ohsuga et al. (2009) and Ohsuga \& Mineshige (2011)}

Last but not least, we mention earlier works by \cite{ohsuga+09} and
\cite{ohsugamineshige-11} who for the first time reproduced three
distinct modes of accretion flows (hot and thick, thin and slim)
within radiation-magnetohydrodynamic simulations. The authors adopted
simplified radiation treatment (flux-limited diffusion), including only free-free and
bound-free opacities, and did not distinguish electron and gas
temperatures. Their setup was sufficient to qualitatively simulate the
three accretion modes. However, it would not be enough for the purpose
of calculating observables.

\subsection{From simulations to observables}

The algorithm presented here allows for precise tracking of gas evolution
and its interaction with the radiation field. The latter is evolved in an
approximate way -- under the grey approximation, imposing a diluted
black body
spectral shape and adopting the M1 closure scheme. Therefore, neither the
local shape of the radiation field (angular distribution of specific
intensities), nor the electromagnetic spectrum is known. From
the simulation output, one can only directly recover the amount of energy
carried by the radiation field and its characteristic temperature.

To obtain the most interesting observable, i.e., the electromagnetic
spectrum for an observer located at a given inclination angle, radiative
postprocessing is necessary. For optically thin flows, i.e., for the
lowest accretion rates, Monte Carlo techniques
\citep[e.g.,][]{roman+12,dexter+10,ck+15a} are most efficient. However, when a
geometrically thin and optically thick disk emerges, a grid-based
approach may be more adequate \citep{narayan+heroic}. One should also take
into account that time-averaged and axisymmetric output is not appropriate for
calculating the spectrum \citep{ck+15a,ck+15b}. Future work in this direction will therefore
require full three-dimensional simulations.

\section{Summary}
\label{s.summary}

In this paper we have introduced a state-of-the-art algorithm which allows
for parallel evolution of gas and radiation field in general
relativity, and which evolves electrons and ions independently, consistently
tracking their temperatures. Such a method allows for the first time
to study the transition from optically thin to optically thick
accretion flows, relevant for many AGN and the hard-to-soft transition of
X-ray binaries.

The new components of the method, when compared with existing
radiation-MHD algorithms, include:

\vspace{.2cm}
(i) Electron and ion energy contents are evolved through their
corresponding entropy conservation equations,  modified by source terms including
viscous heating, Coulomb coupling and radiative heating or cooling.

(ii) The viscous dissipation is distributed between
electrons and ions according to the sub-grid prescription of
\cite{howes+10}.

(iii) The adiabatic indices of each species and their
mixture are adjusted consistently based on their temperatures.

(iv) The radiative spectrum is described as a diluted
black body with, independently evolved, characteristic
temperature.

(v) Bremsstrahlung and synchrotron grey
opacities are obtained by integration over frequencies consistent with the
assumed diluted black body spectral shape. Energy mean
Planck opacities are distinguished from the flux mean Rosseland opacities. The latter
are used in the momentum exchange term.

\vspace{.2cm} We simulated five models of black hole accretion at very
low and moderately low accretion rates to assess the impact and
importance of various factors. We find that radiative effects are
unimportant for gas properties and gas dynamics in the case of a
simulation with accretion rate $\sim 4\times10^{-8}\Medd$. However, in
a model with four orders of magnitude larger accretion rate, the
radiative cooling does become important -- the gas temperature
decreases significantly, and a colder, geometrically thinner structure
forms, though it is still not quite a classic thin disk. The precise
critical accretion rate above which radiative cooling is strong enough
to cause the accretion flow to collapse to a geometrically thin disk
remains to be determined, and will be the subject of a future study.


\section{Acknowledgements}

We thank Sean Ressler, Martin Rees, Chris Fragile and Charles Gammie for comments on the manuscript.
AS acknowledges support
for this work 
by NASA through Einstein Postdoctotral Fellowship number PF4-150126
awarded by the Chandra X-ray Center, which is operated by the
Smithsonian
Astrophysical Observatory for NASA under contract NAS8-03060. AS thanks
Harvard-Smithsonian Center for Astrophysics for its hospitality.
RN was supported in part by NSF grant
AST 1312651 and NASA grant TCAN NNX 14AB7G.
The authors acknowledge computational support from NSF via XSEDE resources
(grant TG-AST080026N),
from NASA via the High-End Computing (HEC) Program
through the NASA Advanced Supercomputing (NAS) Division at Ames
Research Center, and from the PL-Grid Infrastructure. AS acknowledges support from the International Space Science Institute.
This research was supported by the Polish NCN grant UMO-2013/08/A/ST9/00795.
MW acknowledges support of the Foundation for Polish Science within the START programme.
\bibliographystyle{mn2e}

\appendix

\section{Adiabatic Index and Entropy}
\label{ap.entropy}

The pressure $p$ and internal energy $u_{\rm int}$ of a
single-species, non-degenerate, relativistic gas with number density
$n$ and temperature $T$ are given by (e.g., \citealt{chandra39}),
\begin{eqnarray}
p &=& nkT, \label{eq:pressure}\\
u_{\rm int} &=& \rho c^2 \left( \frac{\left[3 K_3(1/\theta) + K_1(1/\theta)\right]}
{4 K_2(1/\theta)} - 1\right) \label{eq:uintexact}\\
&\equiv& \frac{p}{(\gamma_{\rm int} - 1)}, \label{eq:gammaintexact} 
\end{eqnarray}
where $\rho=nm$ is the mass density, $m$ is the particle mass, $\theta
= kT/mc^2$ is the dimensionless temperature, $K_n(x)$ is the modified
Bessel function of order $n$, and we have defined an effective
adiabatic index $\gamma_{\rm int}$ associated with $u_{\rm int}$. The
internal energy has a series expansion of the form,
\begin{equation}
u_{\rm int} = nkT\left(\frac{3}{2} + \frac{15}{8}\theta
-\frac{15}{8}\theta^2\ldots\right), \label{eq:uintseries}
\end{equation}
asymptoting to $(3/2)nkT$ in the non-relativistic limit ($\theta\to0$)
and $3nkT$ in the ultrarelativistic limit
($\theta\to\infty$). Correspondingly, the adiabatic index $\gamma_{\rm
  int}$ asymptotically approaches 5/3 and 4/3 in the two limits.

The specific heat at constant volume $C_V$ and its corresponding
adiabatic index $\gamma_{\rm CV}$ are given by
\begin{eqnarray}
C_V &=& \frac{du_{\rm int}}{dT} \nonumber \\
&=& \frac{nk}{8\theta^2} \left(
\frac{3K_4(1/\theta)+4K_2(1/\theta))+K_0(1/\theta)}
{K_2(1/\theta)}\right. \nonumber \\
&~~~~~~& - \left.\frac{3K_3^2(1/\theta)+4K_3(1/\theta)K_1(1/\theta)
+K_1^2(\theta)}
{K_2^2(1/\theta)}\right) \label{eq:CV} \\
&\equiv& \frac{nk}{\gamma_{\rm CV}-1}. \label{eq:gamma}
\end{eqnarray}
The adiabatic index $\gamma_{\rm CV}$ has a series expansion,
\begin{equation}
\gamma_{\rm CV} = \frac{5}{3} - \frac{5}{3}\theta + \frac{20}{3}\theta^2\cdots,
\label{eq:gammaseries}
\end{equation}
and it again asymptotically approaches to 5/3 and 4/3, respectively, as $\theta\to0$
and $\infty$.  Finally, the specific heat at constant pressure
$C_P$ satisfies
\begin{equation}
C_P = C_V + nk = \gamma_{\rm CV} C_V.
\end{equation}
Above, we have been careful to distinguish between $\gamma_{\rm int}$
and $\gamma_{\rm CV}$. These two effective adiabatic indices are in
general not equal to each other; they match only in the two asymptotic
limits, $\theta\to0$ and $\theta\to\infty$.

While the above expressions are exact, they are not convenient for
numerical computations because of the presence of the Bessel
functions. They are also not easy to ``invert'' to solve for the
temperature, given either the internal energy or the entropy. We
therefore look for simple approximations that are better suited for
computations.

\begin{figure}
 \includegraphics[width=1.\columnwidth]{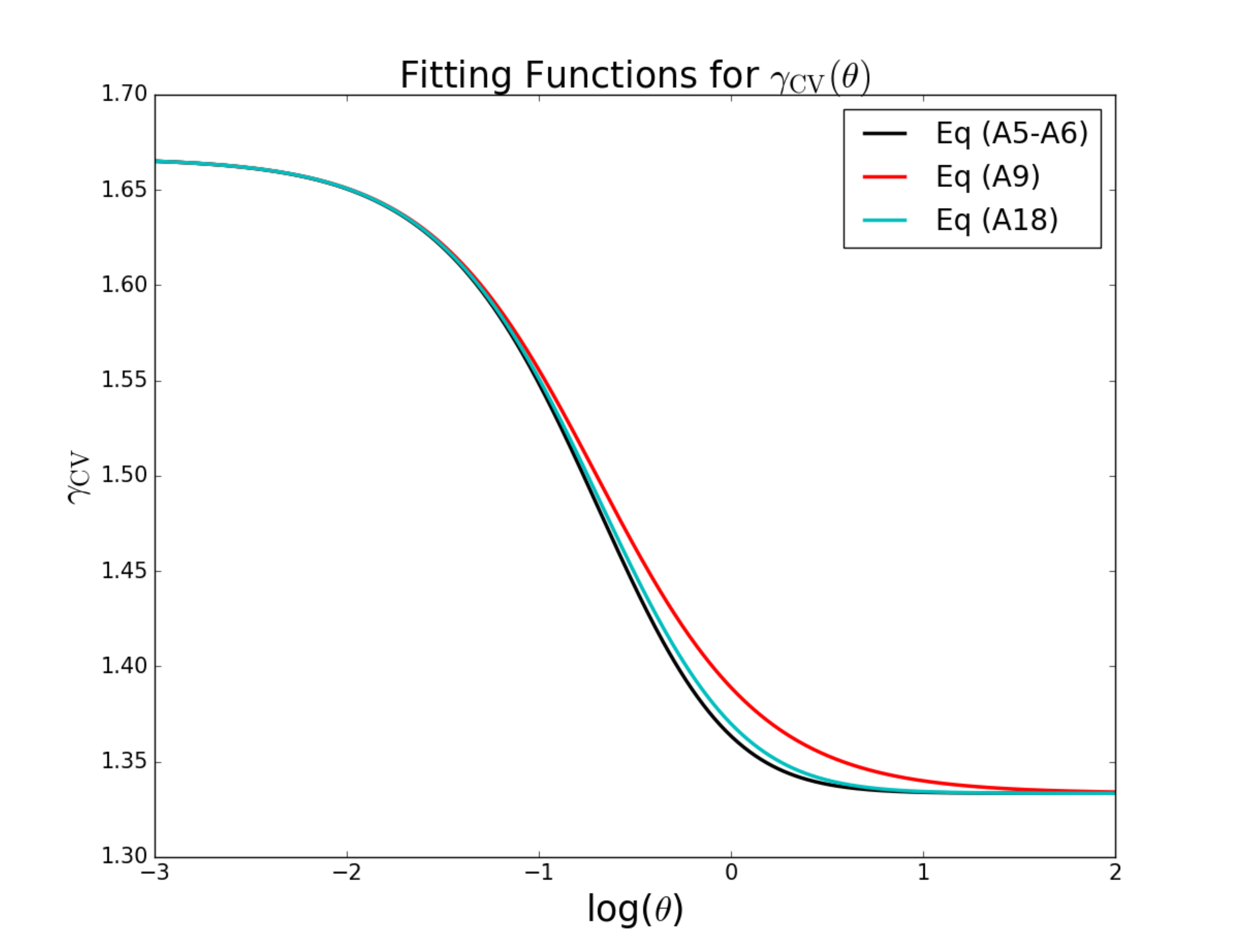}
  \caption{Comparison of the exact solution for $\gamma_{\rm CV}$ as a
    function of the dimensionless temperature $\theta$, obtained from
    equations (\ref{eq:CV}), (\ref{eq:gamma}), with two approximate
    expressions given in equations (\ref{eq:gammafit1}) and
    (\ref{eq:gammafit2}).  We use equation (\ref{eq:gammafit1}) in the
    present work.}
 \label{fig:gamma_fit}
\end{figure}

We begin with $\gamma_{\rm CV}$. Our philosophy is to find a function that
matches the first two terms in the series expansion
(\ref{eq:gammaseries}) and also satisfies the asymptotic result
$\gamma_{\rm CV}\to4/3$ as $\theta\to\infty$. One convenient function is
\begin{equation}
\gamma_{\rm CV} \approx \frac{5+20\theta}{3+15\theta}. \label{eq:gammafit1}
\end{equation}
Figure \ref{fig:gamma_fit} compares this approximation with the exact
result (eqs.~\ref{eq:CV}, \ref{eq:gamma}). We see that the
approximation is very good for small values of $\theta$ up to about
0.1 and also at large $\theta$ above a few tens. In between, the
agreement is not as good, but probably acceptable.  As discussed in the
main text, our simulations generally have non-relativistic ions with
$\theta_i\sim 0.01-0.1$ and relativistic electrons with
$\theta_e>10$. For these temperature ranges, the approximation
(\ref{eq:gammafit1}) is quite good.

We now use equation (\ref{eq:gammafit1}) to derive an approximate
expression for the entropy $s$ per particle:
\begin{equation}
Tds = \frac{C_V}{n}dT + pdv = \frac{k}{(\gamma_{\rm CV}-1)} dT - \frac{kT}{\rho}d\rho,
\end{equation}
where $v$ is the volume per particle.  Using equation
(\ref{eq:gammafit1}) for $\gamma_{\rm CV}$ and integrating, this gives
\begin{equation}
\label{eq.s.ap}
s \approx k \ln \left[ \frac{\theta^{3/2} \left(\theta +
    \frac{2}{5}\right)^{3/2}}{\rho} \right] + C,
\end{equation}
where $C$ is the integration constant. The main advantage of this
approximate formula is that, for given $s$ and $\rho$, it can be
easily inverted to calculate the temperature. Ignoring the integration
constant $C$ (which is equivalent to making a specific choice for the
zero-point of the entropy), we obtain 
\be
\label{e.TfromS3.ap}
\theta \approx \frac{1}{5}\left(\sqrt{1+25 \left[\rho
    \exp\left(\frac{s}{k}\right) \right]^{2/3}}-1\right).  
\ee

We also need an approximation for $u_{\rm int}$. One option is to
substitute (\ref{eq:gammafit1}) in the following equation,
\begin{equation}
\frac{du_{\rm int}}{dT} = C_V = \frac{nk}{\gamma_{\rm CV}-1}, \label{eq:CV2}
\end{equation}
and to integrate. This gives
\begin{equation}
u_{\rm int} \approx \rho c^2\left[3\theta - \frac{3}{5}\ln
\left(1+\frac{5}{2}\theta\right)\right]. \label{eq:uint2}
\end{equation}
Given $u_{\rm int}$ and $\rho$, this transcendental equation can be
solved iteratively for $\theta$; the solution generally converges
quickly. As explained below, we have followed this method in some test
runs.

For computational speed, it is often useful to have a formula for
$u_{\rm int}$ that can be be inverted analytically. For this, we
employ the same philosophy as we used for $\gamma_{\rm CV}$, viz., we
look for a simple approxmation of $u_{\rm int}$ which behaves
correctly in the two asymptotic limits, $\theta\to0$,
$\theta\to\infty$, and also fits the second term in the Taylor series
(\ref{eq:uintseries}).  Such a function is the following,
\begin{equation}
u_{\rm int} \approx \rho c^2 \frac{\theta(6+15\theta)}{(4+5\theta)},
\label{eq:uintapprox}
\end{equation}
which inverts to give,
\begin{equation}
\theta \approx \frac{1}{30}\left(-6+5 \frac{u_{\rm int}}{\rho c^2} + 
\sqrt{36+180 \frac{u_{\rm int}}{\rho c^2}+25 \left(\frac{u_{\rm int}}{\rho c^2}\right)^2}\right).
\label{eq:theta_from_uint}
\end{equation}
The above approximation for $u_{\rm int}$ can be written in terms of
the effective adiabatic index $\gamma_{\rm int}$ (cf,
eq.~\ref{eq:gammaintexact}),
\begin{equation}
\gamma_{\rm int} \approx \frac{10+20\theta}{6+15\theta}, \quad
u_{\rm int} \equiv \frac{p}{\gamma_{\rm int}-1}.
\label{eq:gammaint}
\end{equation}
This form of $\gamma_{\rm int}$ was previously used by
\citet{narayan+gamma}.

For completeness, if we take the approximation (\ref{eq:uintapprox})
for $u_{\rm int}$ and substitute it in (\ref{eq:CV2}), we obtain the
following expression for $\gamma_{\rm CV}$,
\begin{equation}
\gamma_{\rm CV} \approx \frac{20\, (2+8\theta + 5\theta^2)}{3\, (8+40\theta+25\theta^2)}, \label{eq:gammafit2}
\end{equation}
which agrees very well with the exact result for $\gamma_{\rm CV}$
(Fig.~\ref{fig:gamma_fit}). However, this expression involves the
ratio of quadratic factors of $\theta$, and it does not give an
analytically invertible solution for the entropy $s$ (the inversion
again involves a transcendental equation). It is therefore less useful
than equations (\ref{eq:gammafit1}), (\ref{eq.s.ap}),
(\ref{e.TfromS3.ap}), (\ref{eq:theta_from_uint}).

We have run tests in which we used equation (\ref{eq:gammafit1}) for
$\gamma_{\rm CV}$, (\ref{eq.s.ap}) for $s$ and (\ref{e.TfromS3.ap})
for computing $\theta$ from $s$, and tried two different approaches
for inverting $u_{\rm int}$. In the first approach, we numerically
inverted equation (\ref{eq:uint2}) to obtain $\theta$ for a given
value of $u_{\rm int}$, while in the second, we used the approximate
inversion formula (\ref{eq:theta_from_uint}). The two methods gave
virtually identical results. We therefore used the second method in
the simulations discussed in the paper.

We note one final inconsistency: our approximations do not satisfy
Taub's inequality \citep{taub48,mignone07}. However, we believe this
is not a serious concern.

\section{Opacities and emissivities}
\label{ap.opacities}

The frequency dependent opacities $\kappa_\nu$ are related to the frequency dependent emissivities, $\epsilon_\nu$ with the Kirchhoff's law,
\be 
\kappa_\nu (\nu, T_{\rm e}) = \frac{\epsilon_\nu (\nu, T_{\rm e})}{4 \pi \rho \widehat{ B}_\nu(\nu,T_{\rm e})} ,
\label{eq:kirch}
\ee
where $\widehat B_\nu$ is a spectral radiance of a~black body,
\be 
\widehat B_\nu (\nu, T) = \frac{2 h \nu^3}{c^2} \frac{1}{\exp(h \nu / k T) -1} .
\ee
In the notation that we adopt, opacities $\kappa_\nu$ are expressed in units of area per mass. It is often convenient to give formulas for opacity times gas density, which is in units of inverse length. Mean (grey) opacities are obtained by averaging $\kappa_\nu$ over
$\nu$ with proper weights. Clearly, there is no single choice of
weights that would preserve all the physical properties of a~frequency-resolved $\kappa_\nu$ \citep{mihalasbook}. 

In global simulations of accretion flows the Rosseland mean is commonly adopted,
\be 
\kappa_{\rm R} \equiv  \kappa_{\rm R} (T_{\rm e}, T_{\rm r}) = \frac{\int_0^\infty \frac{ \partial \widehat{E}_\nu(\nu , T_{\rm r})}{\partial T_{\rm r}} d \nu}{ \int_0^\infty \kappa_\nu^{-1}(\nu, T_{\rm e}) \frac{ \partial \widehat{E}_\nu(\nu , T_{\rm r})}{\partial T_{\rm r}} d \nu} ,
\label{eq:RossOpacDefin}
\ee
where $\widehat{E}_\nu(\nu , T_{\rm r})$ denotes  the radiation spectral
energy distribution  at a~given time and location. Under our simplifying assumptions we have $\widehat{E}_\nu(\nu , T_{\rm r}) \propto \widehat{B}_\nu(\nu , T_{\rm r}) $, i.e., $\widehat{E}_\nu(\nu , T_{\rm r})$ corresponds to a diluted black body spectrum, see Section \ref{s.gt}.

The Rosseland mean properly addresses conservation of momentum in an
optically thick gas. We employ it exclusively in the momentum
equations by designing a numerical fit to Eq. \ref{eq:RossOpacDefin}. 

The energy equation coupling term, $\widehat G^t$ (Eq.~\ref{eq.Gff2}), consists of emission and absorption terms, for which the energy mean opacities are employed,
\be 
\kappa_{\rm P, e} \equiv \kappa_{\rm P, e}(T_{\rm e}, T_{\rm r}) = \frac{\int_0^\infty \kappa_\nu(\nu, T_{\rm e})  \widehat{B}_\nu(\nu , T_{\rm e}) d \nu}{\int_0^\infty \widehat{B}_\nu(\nu , T_{\rm e}) d \nu} ,
\label{eq:EmiOpacDefin}
\ee
\be 
\kappa_{\rm P, a} \equiv \kappa_{\rm P, a}(T_{\rm e}, T_{\rm r}) = \frac{\int_0^\infty \kappa_\nu(\nu, T_{\rm e})  \widehat{E}_\nu(\nu , T_{\rm r}) d \nu}{\int_0^\infty \widehat{E}_\nu(\nu , T_{\rm r}) d \nu} .
\label{eq:AbsOpacDefin}
\ee
The emission energy mean opacity (Planck opacity) $\kappa_{\rm P, e}$ is weighted with the black body spectral energy density distribution characterized by
the electron temperature, $T_{\mathrm{e}}$. It recovers the total emission from radiating electrons, i.e., 
\be
\epsilon_{\rm tot} \equiv \int_0^\infty \epsilon_\nu (\nu, T_{\rm e}) \, d \nu  =  \kappa_{\rm P, e} \rho \, 4 \pi \widehat{B} .
\ee
The absorption mean opacity is weighted with the radiation spectral
energy distribution  $\widehat{E}_\nu(\nu , T_{\rm r})$, in our case
corresponding to the diluted black body spectrum. Under such an
assumption, this is equivalent to averaging $\kappa_\nu$ with the
Planck distribution, $ \widehat{B}_\nu(\nu , T_{\rm r})$, at a radiation
temperature $T_{\mathrm{r}}$. 
The absorption mean energy opacities depend on both electron and
radiation temperatures.
As $T_{\rm r} \rightarrow T_{\rm e}$, we find that $\kappa_{\rm P, a}
\rightarrow \kappa_{\rm P, e}$.

 In the equation for the photon density evolution, Eq. \ref{eq.cons4} and Eq. \ref{eq:photondens}, we use yet another mean opacity - the number mean opacity $\kappa_{\rm n}$, weighted with the spectral density of the photon number density
\be 
\widehat n_\nu = \frac{\widehat{E}_\nu(\nu , T_{\rm r})}{h \nu} ,
\ee
hence,
\be 
\kappa_{\rm n} \equiv \kappa_{\rm n}(T_{\rm e}, T_{\rm r}) = \frac{\int_0^\infty \kappa_\nu (\nu, T_{\rm e}) \widehat{E}_\nu(\nu , T_{\rm r}) \nu^{-1} d \nu}{\int_0^\infty \widehat{E}_\nu(\nu , T_{\rm r}) \nu^{-1} d \nu} .
\label{eq:NumbOpacDefin}
\ee

We now give the spectrally resolved emmisivity and opacity coefficients for the
bremsstrahlung and synchrotron processes, from which mean opacities
can be calculated using Eqs. \ref{eq:RossOpacDefin} -
\ref{eq:NumbOpacDefin}. For computational reasons the improper integrals that define mean opacities need to be approximated using numerical fits, given in
Eqs. \ref{eq:OpacEnergyAbsFF} - \ref{eq:numopac} as well as further in this appendix.

\subsection{Bremsstrahlung}
\begin{figure}
 \includegraphics[width=1.\columnwidth]{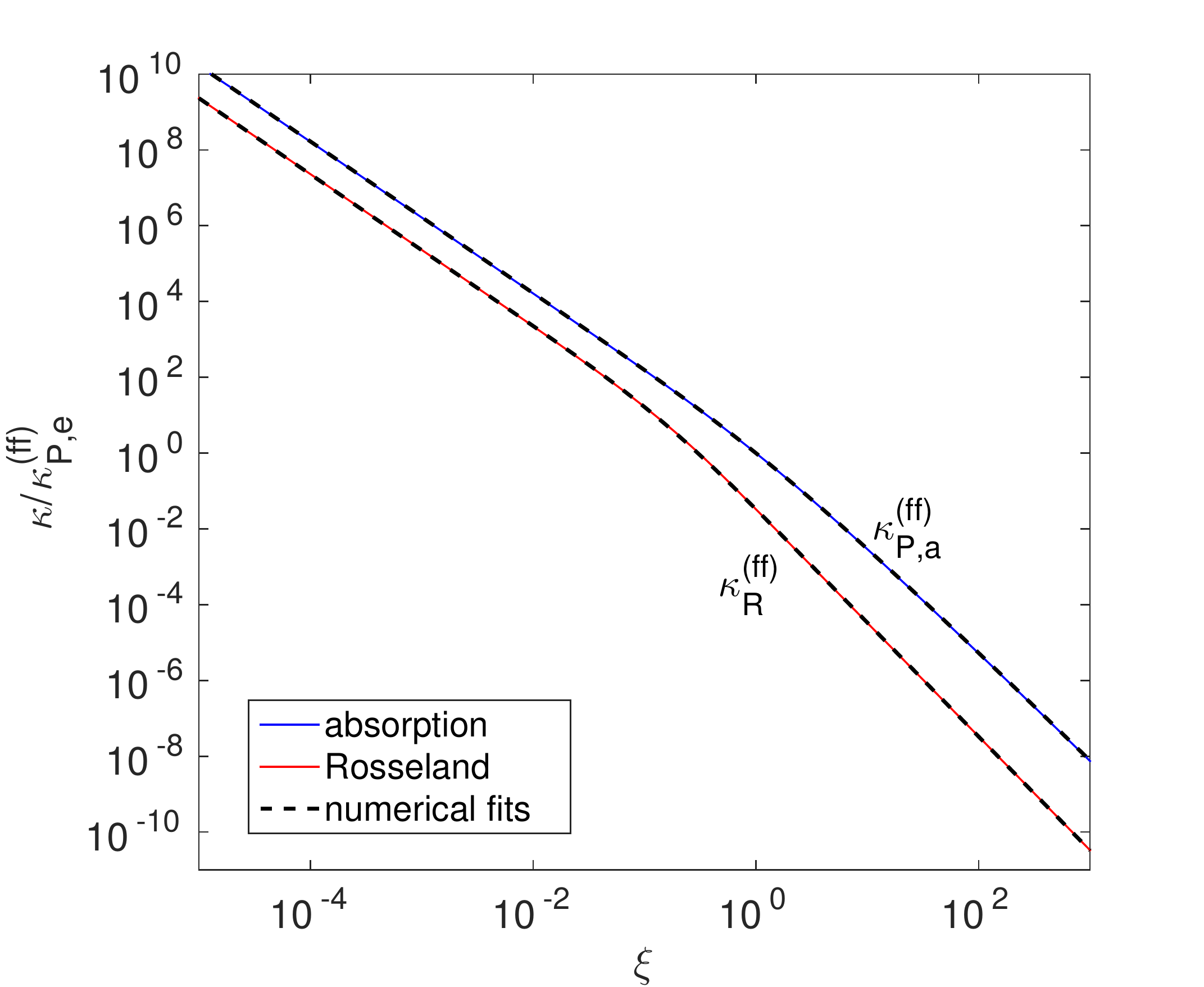}
  \caption{Opacities used for the bremsstrahlung radiation. Error of fits for the shown range of parameter $\xi$ is less than $5\%$. 
  }
 \label{fig:FitsFF}
\end{figure}

Using the notation from Section \ref{sec:opacities}, bremsstrahlung frequency dependent emissivity is equal to \citep{rybicki}
\be 
\epsilon^{ \rm (ff)}_\nu = 6.8 \times 10^{-38} T_{\rm e}^{-1/2} \overline{n} n_{\rm e}   \overline{g} \, R(T_{\rm e}) \exp \left(-h \nu / k T_{\rm e} \right) {\rm \left[ \frac{erg}{cm^3 s \, Hz} \right]},
\ee
from which we find the integrated total emissivity
\be 
\epsilon^{ \rm (ff)} = 1.4 \times 10^{-27} T_{\rm e}^{1/2} \overline{n} n_{\rm e}   \overline{g} \, R(T_{\rm e})\, {\rm \left[ \frac{erg}{cm^3 s} \right]},
\label{eq:totEmisFF}
\ee
The spectrally resolved opacity coefficient is found using Kirchhoff's law,
\be 
\kappa_\nu^{\rm (ff)} \rho = 3.7 \times 10^8 T_{\rm e}^{-1/2} \overline{n} n_{\rm e}   \overline{g} \, R(T_{\rm e}) \frac{1 - \exp (-h \nu / k T_{\rm e})}{\nu^3} {\rm [ cm^{-1} Hz^{-1} ]} .
\label{eq:ff_kappa_nu}
\ee
Inserting  Eq. \ref{eq:ff_kappa_nu} into Eqs. \ref{eq:EmiOpacDefin}-\ref{eq:AbsOpacDefin} and substituting $x = h \nu / k T_{\rm e}$, we find the energy mean opacities
\begin{align}
\label{eq:calcFFemisOpac}
\kappa^{\rm (ff)}_{\rm P, e} \, \rho &= 3.7 \times 10^8  \overline{n} n_{\rm e}   \overline{g} \, R(T_{\rm e}) T_{\rm e}^{-7/2} \left(\frac{h}{k} \right)^3 \frac{I_1 }{I_2 } \nonumber \\
&=   6.2 \times 10^{-24} \overline{n} n_{\rm e}  \,  \overline{g} \, R(T_{\rm e}) T_{\rm e}^{-7/2} , \\
\kappa^{\rm (ff)}_{\rm P, a} \,  \rho &= 3.7 \times 10^8  \overline{n} n_{\rm e}   \overline{g} \, R(T_{\rm e}) T_{\rm e}^{-7/2} \left(\frac{h}{k} \right)^3 \xi^{-4} \frac{I_3(\xi)}{I_2} \nonumber \\
&= \kappa^{\rm (ff)}_{\rm P, e} \,  \rho \xi^{-4} I_3(\xi) \approx \kappa^{\rm (ff)}_{\rm P, e} \,  \rho \xi^{-3} \cdot 1.047 \ln  \left(1 \! + \! 1.6 \xi \right)  .
\label{eq:calcFFabsOpac}
\end{align}
The dimensionless parameter $\xi$ corresponds to temperatures ratio $\xi = T_{\rm r}/T_{\rm e}$. Using Eqs. \ref{eq:totEmisFF} and \ref{eq:calcFFemisOpac} it is easy to verify that $\epsilon^{ \rm (ff)} =  \kappa^{\rm (ff)}_{\rm P, e} \rho 4 \sigma T_{\rm e}^4$. Subsequently, the Rosseland mean opacity is calculated with Eq. \ref{eq:RossOpacDefin},
\begin{align}
\label{eq:calcFFRossOpac}
\kappa^{\rm (ff)}_{\rm R} \, \rho &= 3.7 \times 10^8  \overline{n} n_{\rm e}   \overline{g} \, R(T_{\rm e}) T_{\rm e}^{-7/2} \left(\frac{h}{k} \right)^3 \frac{I_4 }{I_5(\xi) } \nonumber \\
&=   \kappa^{\rm (ff)}_{\rm P, e} \, \rho \xi^{-3} I_2 \frac{I_4}{I_5(\xi)} \approx \kappa^{\rm (ff)}_{\rm P, e} \, \rho \xi^{-3} \cdot 14.12  f_{\rm R}(\xi),
\end{align}
where
\be 
f_{\rm R}(\xi) = \left(432.7 - 106.8 \xi^{-3/5} + 43.17 \xi^{-4/5} + 57.88 \xi^{-1}  \right)^{-1} .
\ee
The accuracy of the numerical fits given by Eqs. \ref{eq:calcFFabsOpac} - \ref{eq:calcFFRossOpac} is indicated in Fig. \ref{fig:FitsFF}. The relevant integrals used in Eqs. \ref{eq:calcFFemisOpac}-\ref{eq:calcFFRossOpac} are
\begin{align}
I_1 &= \int_0^\infty \frac{1-\exp(-x)}{\exp (x) - 1} d x = 1 , \\
I_2 &= \int_0^\infty \frac{x^3}{\exp(x)-1} dx = \frac{\pi^4}{15} , \\
I_3(\xi) &= \int_0^\infty \frac{1-\exp(-x/\xi)}{\exp (x) - 1} d x \approx \xi^{-1} \ln (1 + 1.6 \xi) , \\
I_4 &= \int_0^\infty \frac{x^4 \exp(x)}{[ \exp (x) - 1]^2} d x = \frac{4 \pi^4}{15} , \\
I_5(\xi) &= \int_0^\infty \frac{(x/ \xi)^7 \exp(x/\xi) d (x / \xi) }{[ \exp (x/\xi) - 1]^2[1 - \exp(-x)]}  
\approx \frac{11.95}{ f_{\rm R}(\xi) } .
\end{align}
In the case of the bremsstrahlung radiation, integral \ref{eq:NumbOpacDefin} diverges, indicating that an infinite number of low-energy photons are generated. This divergence can be regularized with a~more detailed treatment of the collective plasma effects, see, e.g., \cite{weldon94}. In this work, for the sake of simplicity, we substitute the photons absorption opacity with the energy absorption mean opacity, 
\be 
\kappa^{\rm (ff)}_{\rm n} \equiv \kappa^{\rm (ff)}_{\rm P, a} .
\ee
We estimate the number of the emitted photons by dividing the total energy produced in the bremsstrahlung process by the (black body) mean emitted photon energy,
\be
\widehat{\dot{n}}_{\rm ff} = \frac{\epsilon^{ \rm (ff)}}{\langle h \nu \rangle } = \frac{\kappa^{\rm (ff)}_{\rm P,e} \, \rho  4 \pi \widehat{B}}{2.7012k T_{\rm e}} .
\ee

\subsection{Synchrotron}

We follow \cite{mahadevan96} and \cite{esin+96} who give the ultrarelativistic gas synchrotron emissivity in Gaussian-cgs units as
\begin{equation}
\epsilon_\nu^{\rm (sy)} = 4.43\times10^{-30} 4\pi \nu_M
n_{\rm e} \,\frac{x_MI'(x_M)}{2\theta_{\rm e}^2} \  {\rm \left[ \frac{erg}{cm^3 s \, Hz} \right]} ,
\label{e.nusyn}
\end{equation}
where
\begin{equation}
x_M = \frac{\nu}{\nu_M}, \qquad \nu_M = \frac{3}{2} \,
\frac{eB}{2\pi m_{\rm e}c}\,\theta_{\rm e}^2 = 1.19 \times 10^{-13} B T_{\rm e}^2 ~{\rm [Hz ]}.
\end{equation}
$I'(x_M)$ is a~fitting function provided by \cite{mahadevan96},
\begin{equation}
I'(x_M) = \frac{4.0505}{x_M^{1/6}} \left(1 + \frac{0.40}{x_M^{1/4}}
+ \frac{0.5316}{x_M^{1/2}} \right) \exp(-1.8899x_M^{1/3}).
\end{equation}
Integrating Eq. \ref{e.nusyn} we find the total emissivity
\be
\epsilon^{\rm (sy)} = 3.61 \times 10^{-34} n_{\rm e} B^2 T_{\rm e}^2 ,
\ee
which allows to find the emission energy mean opacity
\be
\kappa^{\rm (sy)}_{\rm P, e} \, \rho = \frac{\epsilon^{\rm (sy)}}{4 \sigma T_{\rm e}^4} = 1.59 \times 10^{-30} n_{\rm e} B^2 T_{\rm e}^{-2} \ {\rm [ cm^{-1} ] }.
\label{eq:CalcSyEmi}
\ee
We proceed by employing the Rayleigh-Jeans approximation, since synchrotron processes are generally
at very low frequencies ($h\nu \ll kT_{\rm e}$). We then obtain the frequency resolved opacity function $\kappa_\nu$ with the use of the Eq. \ref{eq:kirch},
\begin{equation}
\kappa_\nu^{\rm (sy)} \rho = 2.13 \times 10^{39} n_{\rm e} B^{-1} T_{\rm e}^{-5}
\, \frac{I'(x_M)}{x_M}   {\rm [ cm^{-1} Hz^{-1} ]} .
\end{equation}
We define a~dimensionless parameter $\zeta$
\be 
\zeta = \frac{k T_{\rm r}}{h \nu_M} = 1.745 \times 10^{23} \xi \left[\frac{B}{1 G} \frac{T_{\rm e}}{1 K} \right]^{-1},
\ee
where $\xi = T_{\rm r}/T_{\rm e}$, so that $h \nu / k T_{\rm r} = x_M/ \zeta$. Remaining mean opacities can be found by evaluating integrals \ref{eq:RossOpacDefin}, \ref{eq:AbsOpacDefin} and \ref{eq:NumbOpacDefin},
\begin{align}
\kappa^{\rm (sy)}_{\rm R} \, \rho &= 4.01 \times 10^{-31} n_{\rm e} B^2 T^{-2}_{\rm e} \xi^{-3}  \zeta^{8} \frac{I_4}{I_6(\zeta)} \nonumber \\
&= \kappa^{\rm (sy)}_{\rm P,e}  \, \rho \, \xi^{-3} 6.54 \, \zeta^{8} I^{-1}_6(\zeta) \nonumber \\
&\approx \kappa_{\rm P, e}^{\rm (sy)} \, \rho \, \xi^{-3} \cdot  3.24 \times 10^{-2} \zeta^{1.31} \exp \left(-1.60\zeta^{0.463} \right) ,\label{eq:CalcSyRos} \\
\kappa^{\rm (sy)}_{\rm P,a}  \, \rho &= 4.01 \times 10^{-31} n_{\rm e} B^2 T^{-2}_{\rm e} \xi^{-3} \zeta^{-1} \frac{I_7(\zeta)}{I_2} \nonumber \\
&= \kappa^{\rm (sy)}_{\rm P,e}  \, \rho \, \xi^{-3} 3.88 \times 10^{-2} \zeta^{-1} I_7(\zeta) \nonumber \\
&\approx \kappa_{\rm P, e}^{\rm (sy)} \, \rho \, \xi^{-3} \left(1 + 5.444 \zeta^{-2/3} + 7.218 \zeta^{-4/3} \right)^{-1} ,\label{eq:CalcSyAbs}\\
\kappa^{\rm (sy)}_{\rm n}  \, \rho &= 4.01 \times 10^{-31} n_{\rm e} B^2 T^{-2}_{\rm e} \xi^{-3} \frac{I_8(\zeta)}{I_9} \nonumber \\
&= \kappa^{\rm (sy)}_{\rm P,e}  \, \rho \, \xi^{-3} 1.05 \times 10^{-1} I_8(\zeta) \nonumber \\
&\approx \kappa_{\rm P, e}^{\rm (sy)} \, \rho \, \xi^{-3} \cdot 0.868 \zeta \left(1 +0.589 \zeta^{-1/3} + 0.087 \zeta^{-2/3} \right)^{-1}. \label{eq:CalcSyNum}
\end{align}
where the relevant integrals are
\begin{align}
I_6(\zeta) &= \int_0^\infty  \frac{ x_M^5  \exp (x_M/\zeta)d x_M }{I' (x_M)[ \exp (x_M/\zeta) - 1]^2} \nonumber \\ 
&\approx 202 \zeta^{6.69} \exp \left(1.60 \zeta^{0.463} \right) , \\
I_7(\zeta) &= \int_0^\infty  \frac{I' (x_M) x_M^2 d x_M }{[ \exp (x_M/\zeta) - 1]}  \nonumber \\ 
&\approx \frac{26.0 \zeta}{1 + 5.444 \zeta^{-2/3} + 7.218 \zeta^{-4/3}} , \\
I_8(\zeta) &= \int_0^\infty  \frac{I' (x_M) x_M d x_M }{[ \exp (x_M/\zeta) - 1]}  \nonumber \\ 
&\approx \frac{8.27 \zeta}{1 + 0.589 \zeta^{-1/3} + 0.087 \zeta^{-2/3}} , \\
I_9 &= \int_0^\infty  \frac{ x^2 d x }{ \exp (x) - 1}  
\approx 2.404 .
\end{align}
\begin{figure}
 \includegraphics[width=1.\columnwidth]{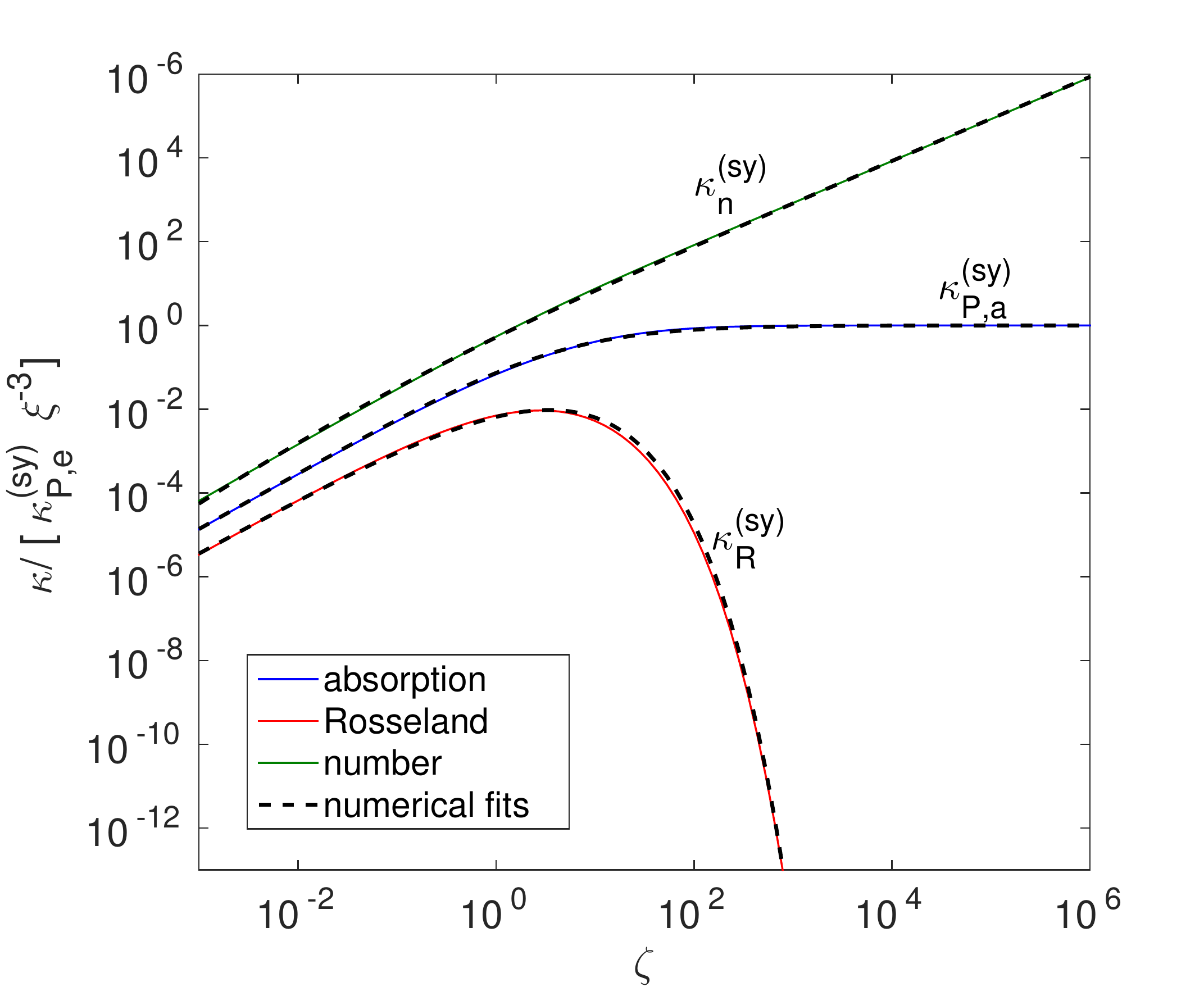}
  \caption{Opacities used for the synchrotron radiation. Error of fits for the shown range of parameter $\zeta$ is less than $15\%$ for $\kappa^{\rm (sy)}_{\rm n}$ and $\kappa^{\rm (sy)}_{\rm P, a}$. The fitting error is slightly larger in case of $\kappa^{\rm (sy)}_{\rm R}$, because of the high variability of this distribution. However, this mostly affects the region of large $\zeta$, where Rosseland opacity becomes negligible anyway.
  }
 \label{fig:FitsSynch}
\end{figure}
The accuracy of the fitting formulas given by Eqs. \ref{eq:CalcSyRos} - \ref{eq:CalcSyNum} is indicated in Fig. \ref{fig:FitsSynch}. Finally, we evaluate the rate at which photons are produced by the synchrotron process by direct integration,
\be
\widehat{\dot{n}}_{\rm sy} = \int_0^\infty \frac{ \epsilon_\nu^{\rm (sy)} }{h \nu} d \nu  = 1.44 \times 10^{5} B n_{\rm e} \ {\rm [cm^{-3} s^{-1}]}.
\ee

\end{document}